\documentclass[journal=jctcce,manuscript=article,articletitle=true,layout=twocolumn]{achemso} 
\setkeys{acs}{articletitle=true}
\setkeys{acs}{doi = false} 

\usepackage[version=4]{mhchem} 
\usepackage{amssymb,amsmath}
\usepackage{tabularx}
\usepackage[center]{subfigure}
\usepackage{multirow}
\usepackage{color}
\usepackage{comment}
\usepackage[usenames,dvipsnames]{xcolor}
\usepackage{graphicx}
\usepackage{refstyle}
\usepackage[colorlinks=true, allcolors=blue]{hyperref}
\usepackage{threeparttable}
\usepackage{attachfile}

\newcolumntype{C}{>{\centering\arraybackslash}X}
\newcolumntype{R}{>{\raggedleft\arraybackslash}X}
\newcolumntype{L}{>{\raggedright\arraybackslash}X}

\newcommand \commentout[1] {}

\newcommand*\citeref[1]{ref.~\citenum{#1}}
\newcommand*\citerefs[1]{refs.~\citenum{#1}}
\newcommand*\jouletesla[1]{\ensuremath{#1 \times 10^{-30}\, {\rm J}/{\rm T}^2}}
\newcommand*{\bohr}[1]{\mbox{\ensuremath{#1 \, a_\mathrm{0}}}}
\newcommand*\nATbohr[1]{\ensuremath{#1 \, {\rm nAT^{-1} }a_0^{-2}}}

\newcommand*\rankst[1]{\ensuremath{#1^\mathrm{st}}}
\newcommand*\ranknd[1]{\ensuremath{#1^\mathrm{nd}}}
\newcommand*\rankrd[1]{\ensuremath{#1^\mathrm{rd}}}
\newcommand*\rankth[1]{\ensuremath{#1^\mathrm{th}}}

\newcommand*\ie{\emph{i.e.}}

\newcommand*\Gimic{\textsc{Gimic}}
\newcommand*\Numgrid{\textsc{Numgrid}}
\newcommand*\PySCF{\textsc{PySCF}}
\newcommand*\Gaussian{\textsc{Gaussian}}
\newcommand*\Turbomole{\textsc{Turbomole}}
\newcommand*\Libxc{\textsc{Libxc}}
\newcommand*\XCFun{\textsc{XCFun}}

\SectionNumbersOn

\author{Susi Lehtola}
\email{susi.lehtola@alumni.helsinki.fi}
\affiliation{University of Helsinki, Department of Chemistry, P.O. Box
  55 (A.I. Virtanens plats 1), FI-00014 University of Helsinki, Finland}
\alsoaffiliation{Molecular Sciences Software
  Institute, Blacksburg, Virginia 24061, United States}

\author{Maria Dimitrova}
\email{maria.dimitrova@helsinki.fi}
\affiliation{University of Helsinki, Department of Chemistry, P.O. Box
  55 (A.I. Virtanens plats 1), FI-00014 University of Helsinki, Finland}

\author{Heike Fliegl}
\email{heike.fliegl@kit.edu}
\affiliation{KIT, Institute of Nanotechnology, Hermann-von-Helmholtz Platz 1, D-76344 Eggenstein-Leopoldshafen, Germany}
 
\author{Dage Sundholm}
\email{dage.sundholm@helsinki.fi}
\affiliation{University of Helsinki, Department of Chemistry, P.O. Box
  55 (A.I. Virtanens plats 1), FI-00014 University of Helsinki, Finland}

\title{Benchmarking magnetizabilities with recent density functionals}

\abbreviations{NMR,GIMIC}

\begin{document}



\begin{abstract}
We have assessed the accuracy for magnetic properties of a set of 51
density functional approximations, including both recently published
as well as already established functionals. The accuracy assessment
considers a series of 27 small molecules and is based on comparing the
predicted magnetizabilities to literature reference values calculated
using coupled cluster theory with full singles and doubles and
perturbative triples [CCSD(T)] employing large basis sets. The most
accurate magnetizabilities, defined as the smallest mean absolute
error, were obtained with the BHandHLYP functional. Three of the six
studied Berkeley functionals and the three range-separated Florida
functionals also yield accurate magnetizabilities. Also some older
functionals like CAM-B3LYP, KT1, BHLYP (BHandH), B3LYP and PBE0
perform rather well. In contrast, unsatisfactory performance was
generally obtained with Minnesota functionals, which are therefore not
recommended for calculations of magnetically induced current
density susceptibilities, and related magnetic properties such as
magnetizabilities and nuclear magnetic shieldings.

We also demonstrate that magnetizabilities can be calculated by
numerical integration of the magnetizability density; we have
implemented this approach as a new feature in the gauge-including
magnetically induced current method (\Gimic{}). Magnetizabilities can
be calculated from magnetically induced current density
susceptibilities within this approach even when analytical approaches
for magnetizabilities as the second derivative of the energy have not
been implemented. The magnetizability density can also be visualized,
providing additional information that is not otherwise easily
accessible on the spatial origin of the
magnetizabilities.  \end{abstract}

\section{Introduction}
\label{sec:intro}

Computational methods based on density-functional theory (DFT) are
commonly used in quantum chemistry, because DFT calculations are
rather accurate despite their relatively modest computational
costs. Older functionals such as the
Becke'88--Perdew'86\cite{Becke1988_3098, Perdew1986_8822} (BP86),
Becke'88--Lee--Yang--Parr\cite{Becke1988_3098, Lee1988_785} (BLYP) and
Perdew--Burke--Ernzerhof\cite{Perdew1996_3865, Perdew1996_3865_err}
(PBE) functionals at the generalized gradient approximation (GGA) as
well as the B3LYP\cite{Stephens1994_11623} and
PBE0\cite{Adamo1999_6158, Ernzerhof1999_5029} hybrid functionals are
still often employed, even though newer functionals with improved
accuracy for energies and electronic properties have been developed.

The accuracy and reliability of various density functional
approximations (DFAs) has been assessed in a huge number of
applications and benchmark studies.\cite{SilvaJunior2008_JCP_104103,
  Sauer2009_JCTC_555, SilvaJunior2010_JCP_174318,
  Laurent2013_IJQC_2019, Mardirossian2016_JCTC_4303,
  Goerigk2011_PCCP_6670, Mardirossian2017_MP_2315,
  Stoychev2018_JCTC_619, Grabarek2019_JCTC_490} It is important to
note that functionals that are accurate for energetics may be less
suited for calculations of other molecular
properties.\cite{Stoychev2018_JCTC_619} In specific, the accuracy of
magnetic properties calculated within DFAs has been benchmarked by
comparing magnetizabilities and nuclear magnetic shieldings to those
obtained from coupled-cluster calculations using large basis
sets,\cite{Lutnes2009_JCP_144104, Teale2013_JCP_24111} although modern
DFAs have been less systematically
investigated.\cite{Zhao2008_JPCA_6794, Johansson2010_JCTC_3302,
  Gromov2019_JMM_93, ZunigaGutierrez2012_JCP_94113,
  Stoychev2018_JCTC_619} The same also holds for nuclear independent
chemical shifts\cite{Chen2005_CR_3842, Sola2010_S_1156,
  Rosenberg2014_CR_5379, GershoniPoranne2015_CSR_6597,
  Gajda2018_MRC_265} and magnetically induced current density
susceptibilities,\cite{Sambe1973_JCP_555, Lazzeretti2000_PNMRS_1,
  Lazzeretti2018_JCP_134109, Juselius2004_JCP_3952,
  Taubert2011_JCP_54123, Fliegl2011_PCCP_20500,
  Sundholm2016_WIRCMS_639, Fliegl2018_CM_1} which have been studied
for a large number of molecules, but whose accuracy has never been
benchmarked properly.

Magnetizabilities are usually calculated as the second derivative of
the electronic energy with respect to the external magnetic
perturbation,\cite{Ruud1993_JCP_3847, Ruud1994_JACS_10135,
  Ruud1995_CP_157, Loibl2014_JCP_24108, Helgaker2012_CR_543}
\begin{equation}
  \xi_{\alpha\beta} = -\frac{\partial^2 E}{\partial B_{\alpha}
  \partial  B_{\beta}}\Bigg|_{{\bf B}={\bf 0}}.
  \label{eq:zeta-secondder}
\end{equation}
Such analytic implementations for magnetizabilities exist in several
quantum chemistry programs.  However, since the magnetic interaction
energy in \eqref{Emag} can also be written as an integral over the
magnetic interaction energy density $\rho^{\bf B}({\bf r})$ that is
given by the scalar product of the magnetically induced current
density $\mathbf{J^B}(\mathbf{r})$ with the vector potential
$\mathbf{A^B}(\mathbf{r})$ of the external magnetic field
$\mathbf{B}$\cite{Jameson1979_JPC_3366, Jameson1980_JCP_5684,
  Fowler1998_JPCA_7297, Ilias2013_MP_1373, Lazzeretti2000_PNMRS_1,
  Lazzeretti2018_JCP_134109}
\begin{equation}
  E = \int \rho^{\bf B}({\bf r}) ~\mathrm{d}^3 r = -\frac{1}{2}\int
  \mathbf{A}^{\mathbf{B}}(\mathbf{r}) \cdot \mathbf{J^B}(\mathbf{r}) ~\mathrm{d}^3r,
  \label{eq:Emag}
\end{equation}
an approach based on quadrature is also possible.  As will be seen in
\secref{theory}, the numerical integration approach for the
magnetizability provides additional information about its spatial
origin that is not available with the analytic approach based on
second derivatives: the tensor components of the magnetizability
density defined in \secref{theory} are scalar functions that can be
visualized, and the integration approach can be used to provide
detailed information about the origin of the corresponding components
of the magnetizability tensor. Similar approaches have been used in
the literature for studying spatial contributions to nuclear magnetic
shielding constants.\cite{Steiner2004_PCCP_261, Pelloni2004_OL_4451,
  Ferraro2004_CPL_268, Soncini2005_CPL_164, Ferraro2005_MRC_316,
  Acke2018_JCC_511, Acke2019_PCCP_3145, Jinger:20}

We will describe our methods for numerical integration of
magnetizabilities using the current density susceptibility in
\secref{theory, implementation}. Then, in \secref{methods}, we will
list the studied set of density functionals, and present the results
in \secref{results}: the functional benchmark is discussed in
\secref{benchmark}, and magnetizability densities and spatial
contributions to magnetizabilities are analyzed in
\secref{magnetizability}. The conclusions of the study are summarized
in \secref{conclusions}. Atomic units are used throughout the text,
unless stated otherwise, and summation over repeated indices is
assumed.

\section{Theory}
\label{sec:theory}

The current density $\mathbf{J^B}(\mathbf{r})$ in \eqref{Emag} is
formally defined as the real part $(\mathcal{R})$ of the mechanical
momentum density,
\begin{equation}
\mathbf{J}^\mathbf{B}(\mathbf{r}) = -\mathcal{R}
\left[\Psi^*(\mathbf{r})
  \left(\mathbf{p}-\mathbf{A}^\mathbf{B}(\mathbf{r})\right)
  \Psi(\mathbf{r})\right],
\label{eq:curdens}
\end{equation}
where $\mathbf{p}=-\mathrm{i}\nabla$ is the momentum
operator. Substituting  \eqref{Emag} into \eqref{zeta-secondder}
straightforwardly leads to
\begin{equation}
  \xi_{\alpha\beta} = \frac{\partial^2}{\partial B_\alpha \partial
    B_\beta} \frac{1}{2}\int \mathbf{A}^{\mathbf{B}}(\mathbf{r}) \cdot
  \mathbf{J^B}(\mathbf{r}) ~\mathrm{d}^3r
\Bigg|_{{\bf B}={\bf 0}}.
  \label{eq:magnetizability}
\end{equation}
The current density susceptibility tensor\cite{Sambe1973_JCP_555,
  Lazzeretti2000_PNMRS_1, Lazzeretti2018_JCP_134109} (CDT) is defined
as the first derivative of the magnetically induced current density
with respect to the components of the external magnetic field in the
limit of a vanishing magnetic field,\cite{Juselius2004_JCP_3952,
  Taubert2011_JCP_54123, Fliegl2011_PCCP_20500,
  Sundholm2016_WIRCMS_639}
\begin{equation}
  {\cal J}^{B_\beta}_{\gamma} = \frac {\partial J_\gamma^{\bf B}} {\partial B_\beta} \Bigg|_{{\bf B}={\bf 0}}.
  \label{eq:MCDSS}
\end{equation}
The vector potential $\mathbf{A}^\mathbf{B}(\mathbf{r})$ of an external
static homogeneous magnetic field is expressed as
\begin{equation}
  \mathbf{A}^\mathbf{B}(\mathbf{r}) = \frac{1}{2} \mathbf{B} \times (\mathbf{r}-\mathbf{R}_O),
  \label{eq:vecpot}
\end{equation}
where $\mathbf{R}_O$ is the chosen gauge origin. The $\alpha \beta$
component of the magnetizability tensor can then be obtained from
\eqref{magnetizability, MCDSS, vecpot} as
\begin{equation}
\xi_{\alpha\beta} = \int \rho_{\alpha \beta}^\xi (\mathbf{r}) \mathrm{d}^3 r,
\label{eq:Biot-Savart}
\end{equation}
where the magnetizability density is defined as
\begin{equation}
\rho_{\alpha\beta}^\xi ({\bf r}) = \frac{1}{2} \sum_{\delta \gamma}
\epsilon_{\alpha\delta\gamma} {r}_\delta {\cal J}^{B_\beta}_{\gamma}
(\mathbf{r})
\label{eq:magdens}
\end{equation}
where $\epsilon_{\alpha \delta \gamma}$ is the Levi--Civita symbol,
$\alpha$, $\beta$, $\gamma$, and $\delta$ are one of the Cartesian
directions $(x,y,z)$, and $r_\delta$ also denotes one of $(x,y,z)$.
The components of the magnetizability density tensor $\rho^\xi_{\alpha
  \beta} ({\bf r})$ are scalar functions that can be visualized to
obtain information about the spatial contributions to the
corresponding element of the magnetizability tensor $\xi_{\alpha
  \beta}$.

As the isotropic magnetizability ($\overline{\xi}$) is obtained as the
average of the diagonal elements of the magnetizability tensor
\begin{equation}
  \overline{\xi}=\frac{1}{3} \mathrm{Tr}~\xi =
  \int \overline{\rho}^\xi({\bf r}) \mathrm{d}^3 r,
  \label{eq:xiave}
\end{equation}
we introduce the isotropic magnetizability density
$\rho^{\overline{\xi}}({\bf r})$ defined as
\begin{equation}
\overline{\rho}^\xi ({\bf r}) = \frac{1}{3} \text{Tr}~\boldsymbol{\rho}^\xi({\bf r}),
\label{eq:isomagdens}
\end{equation}
which yields information about the spatial origin of the isotropic
magnetizability, as we will demonstrate in \secref{magnetizability}.

Although there is freedom with regard to the choice of the gauge
origin of $\mathbf{A}^\mathbf{B}(\mathbf{r})$, the magnetic flux
density $\mathbf{B}$ is uniquely defined via \eqref{vecpot}, because
$\mathbf{B}= \nabla \times (\mathbf{A}(\mathbf{r}) + \nabla
f(\mathbf{r}))$ holds for any differentiable scalar function
$f(\mathbf{r})$.  The exact solution of the Schr{\"o}dinger equation
should also be gauge invariant.  However, the use of finite
one-particle basis sets introduces gauge dependence in quantum
chemical calculations of magnetic properties. The CDT can be made
gauge-origin independent by using gauge-including atomic orbitals
(GIAOs), also known as London atomic orbitals
(LAOs),\cite{Ditchfield1974_MP_789, Wolinski1990_JACS_8251,
  Juselius2004_JCP_3952}
\begin{equation}
    \chi_\mu(\mathbf{r}) = e^{-\mathrm{i} (\mathbf{B}\times [\mathbf{R}_\mu-\mathbf{R}_O] \cdot \mathbf{r})/2} \chi_\mu^{(0)} \left(\mathbf{r}\right), \label{eq:lao}
\end{equation}
where $\mathrm{i}$ is the imaginary unit and
$\chi_\mu^{(0)}(\mathbf{r})$ is a standard atomic-orbital basis
function centered at $\mathbf{R}_\mu$. GIAOs eliminate the gauge
origin from the expression used for calculating the CDT; the
expression we use is given in the supporting information (SI).  Since
the expression for the magnetizability density in \eqref{Biot-Savart,
  magdens} can be computed by quadrature, magnetizabilities can be
obtained from the CDT even if the corresponding analytical
calculation of magnetizabilities as the second derivative of the
energy has not been implemented.

\section{Implementation \label{sec:implementation}}

The present implementation is based on the \Gimic{}
program\cite{gimic-download} and the \Numgrid{} library,\cite{Bast:20}
which are both freely available open-source software.
Gauge-independent CDTs can be calculated with
\Gimic{}\cite{Juselius2004_JCP_3952, Taubert2011_JCP_54123,
  Fliegl2011_PCCP_20500, Sundholm2016_WIRCMS_639} using the density
matrix, the magnetically perturbed density matrices and information
about the basis set.

In order to evaluate \eqref{Biot-Savart}, a molecular integration grid
is first generated from atom-centered grids with the \Numgrid{}
library, as described by \citet{Becke1988_JCP_2547}. In \Numgrid{},
the grid weights are scaled according to the Becke partitioning scheme
using a Becke hardness of 3;\cite{Becke1988_JCP_2547} the
atom-centered grids are determined by a radial grid generated as
suggested by \citet{Lindh2001_TCA_178}, and angular grids due to
\citet{Lebedev1995_RASDM_283} are used.

Given the quadrature grid, the diagonal elements of the
magnetizability tensor are calculated in \Gimic{} from the Cartesian
coordinates of the $n$ grid points multiplied with the CDT calculated in
the grid points. For example, the $\xi_{xx}$ element of the
magnetizability tensor is obtained from \eqref{Biot-Savart} as
\begin{equation}
\xi_{xx} = \sum_{i = 1}^{n} \rho^\xi_{i;xx}
\label{eq:quadrature}
\end{equation}
where the $xx$ component of the magnetizability density tensor at grid
point $i$ is
\begin{equation}
\rho^\xi_{i;xx} = \frac{1}{2} \left [ \left(y {\cal
    J}^{B_x}_{z}\right)_i - \left(z {\cal J}^{B_x}_{y}\right)_i \right
]
\label{eq:Biot-Savart-num}
\end{equation}
where $\left(y {\cal J}^{B_x}_{z}\right)_i$ and $\left(z {\cal
  J}^{B_x}_{y}\right)_i$ are the product of the $z$ and $y$ components
of the CDT calculated in grid point $i$ with the Cartesian
coordinates $y$ and $z$ of the grid point, respectively, and the
external magnetic field perturbation is along the $x$ axis, $B_x$. The
$\xi_{yy}$ and $\xi_{zz}$ elements are obtained analogously.

\section{Computational Methods \label{sec:methods}}

Calculations are performed for the set of 28 molecules studied in
\citeref{Lutnes2009_JCP_144104} that also provides our molecular
structures and the CCSD(T) reference values: \ce{AlF}, \ce{C2H4},
\ce{C3H4}, \ce{CH2O}, \ce{CH3F}, \ce{CH4}, \ce{CO}, \ce{FCCH},
\ce{FCN}, \ce{H2C2O}, \ce{H2O}, \ce{H2S}, \ce{H4C2O}, \ce{HCN},
\ce{HCP}, \ce{HF}, \ce{HFCO}, \ce{HOF}, \ce{LiF}, \ce{LiH}, \ce{N2},
\ce{N2O}, \ce{NH3}, \ce{O3}, \ce{OCS}, \ce{OF2}, \ce{PN}, and
\ce{SO2}.  However, as in \citeref{Lutnes2009_JCP_144104}, \ce{O3} was
omitted from the analysis, since it is an outlier, and due to the fact
that the reliability of the CCSD(T) level of theory is not guaranteed
for this system: the perturbative triples correction to the
magnetizability of \ce{O3} is \jouletesla{-46.2}, indicating that the
CCSD(T) result might still have large error
bars.\cite{Lutnes2009_JCP_144104} The results of this work thus only
pertain to the 27 other molecules, as in
\citeref{Lutnes2009_JCP_144104}.

\begin{table*}
  \caption{Functionals at the local density approximation (LDA) and
    the generalized gradient approximation (GGA) considered in this
    work. GH stands for global hybrid and RS for range separated
    hybrid.  The amount of Hartree--Fock (HF) exchange, or exact
    exchange in the short range (SR) and long range (LR) are also
    given.}
  \begin{threeparttable}
    \begin{tabular}{lccccc}
      \hline
      Functional & Hybrid & Type & Notes & \Libxc{} ID$^a$ & References \\
      \hline
      \hline
      LDA  & & LDA & & 1+7 & \citenum{Bloch1929_545, Dirac1930_376, Vosko1980_1200} \\
      BLYP & & GGA & & 106+131 & \citenum{Becke1988_3098, Lee1988_785, Miehlich1989_200} \\
      BP86 & & GGA & & 106+132 & \citenum{Becke1988_3098, Perdew1986_8822} \\
      CHACHIYO & & GGA & & 298+309 & \citenum{Chachiyo2020_3485, Chachiyo2020_112669} \\
      KT1 & & GGA & & 167 & \citenum{Keal2003_3015} \\
      KT2 & & GGA & & 146 & \citenum{Keal2003_3015} \\
      KT3 & & GGA & \PySCF{} data used & 587 & \citenum{Keal2004_5654} \\
      N12 & & GGA & & 82+80 & \citenum{Peverati2012_2310} \\
      PBE & & GGA & & 101+130 & \citenum{Perdew1996_3865,  Perdew1996_3865_err} \\
      B3LYP & GH & GGA & 20\% HF & 402 & \citenum{Stephens1994_11623} \\
      revB3LYP$^b$ & GH & GGA & 20\% HF & 454 & \citenum{Lu2013_64} \\
      B97-2 & GH & GGA & 21\% HF & 410 & \citenum{Wilson2001_9233} \\
      B97-3 & GH & GGA & 26.9\% HF & 414 & \citenum{Keal2005_121103} \\
      BHLYP$^c$ & GH & GGA & 50\% HF & 435 & \citenum{Bloch1929_545, Dirac1930_376, Becke1993_1372} \\
      BHandHLYP$^d$ & GH & GGA & 50\% HF & 436 & \citenum{Becke1988_3098, Becke1993_1372} \\
      PBE0 & GH & GGA & 25\% HF & 406 & \citenum{Adamo1999_6158, Ernzerhof1999_5029} \\
      QTP17 & GH & GGA & 62\% HF & 416 & \citenum{Jin2018_064111} \\
      N12-SX & RS & GGA & 25\% SR, 0\% LR & 81+79 & \citenum{Peverati2012_16187} \\
      CAM-B3LYP & RS & GGA & 19\% SR, 65\% LR & 433 & \citenum{Yanai2004_51} \\
      CAMh-B3LYP$^e$ & RS & GGA & 19\% SR, 50\% LR & -- & \citenum{Shao2020_587} \\
      CAM-QTP-00 & RS & GGA & 54\% SR, 91\% LR & 490 & \citenum{Verma2014_18A534} \\
      CAM-QTP-01 & RS & GGA & 23\% SR, 100\% LR & 482 & \citenum{Jin2016_034107} \\
      CAM-QTP-02 & RS & GGA & 28\% SR, 100\% LR & 491 & \citenum{Haiduke2018_184106} \\
      $\omega$B97 & RS & GGA & 0\% SR, 100\% LR & 463 & \citenum{Chai2008_084106} \\
      $\omega$B97X & RS & GGA & 15.8\% SR, 100\% LR & 464 & \citenum{Chai2008_084106} \\
      $\omega$B97X-D & RS & GGA & 22.2\% SR, 100\% LR & 471 & \citenum{Chai2008_6615} \\
      $\omega$B97X-V & RS & GGA & 16.7\% SR, 100\% LR & 531 & \citenum{Mardirossian2014_9904} \\
      \hline
    \end{tabular}
    \begin{tablenotes}
    \item [$^a$] Two numbers indicate the exchange and the correlation functional respectively. A single number indicates an exchange-correlation functional.
    \item [$^b$] Revised version
    \item [$^c$] Following King el al. in \citerefs{King1996_JPC_6061,
      King1996_JCP_6880, King1997_JCP_8536}, BHLYP is defined as 50\%
      LDA exchange, 50\% of HF exchange, and 100\% LYP correlation. It
      is sometimes also known as BHandH, which is its keyword in
      \Gaussian{}.
    \item [$^d$] BHandHLYP is 50\% Becke'88 exchange, 50\% HF exchange, and 100\% LYP correlation.
    \item [$^e$] CAMh-B3LYP is defined using the \XCFun{} library with $\alpha = 0.19; \beta = 0.31; \mu = 0.33$.
    \end{tablenotes}
  \end{threeparttable}
  \label{tab:ldas-ggas}
\end{table*}

\begin{table*}
  \caption{Meta-GGA functionals (mGGA) considered in this work. The
    notation is the same as in \tabref{ldas-ggas}.}
  \begin{threeparttable}
    \begin{tabular}{lccccc}
      \hline
      Functional & Hybrid & Type & Notes & \Libxc{} ID$^a$ & References \\
      \hline
      \hline
      B97M-V & & mGGA & & 254 & \citenum{Mardirossian2015_074111} \\
      M06-L & & mGGA & & 449+235 & \citenum{Zhao2006_194101} \\
      revM06-L$^b$ & & mGGA & & 293+294 & \citenum{Wang2017_8487} \\
      M11-L & & mGGA & & 226+75 & \citenum{Peverati2012_117} \\
      MN12-L & & mGGA & & 227+74 & \citenum{Peverati2012_13171} \\
      MN15-L & & mGGA & & 268+269 & \citenum{Yu2016_1280} \\
      TASK & & mGGA & & 707+13 & \citenum{Aschebrock2019_033082, Perdew1992_13244} \\
      MVS & & mGGA & & 257+83 & \citenum{Sun2015_685, Perdew2009_026403} \\
      SCAN & & mGGA & & 263+267 & \citenum{Sun2015_036402} \\
      rSCAN$^c$ & & mGGA & & 493+494 & \citenum{Bartok2019_161101} \\
      TPSS & & mGGA & & 457 & \citenum{Tao2003_146401, Perdew2004_6898} \\
      revTPSS$^b$ & & mGGA & & 212+241 & \citenum{Perdew2009_026403, Perdew2009_026403_err} \\
      TPSSh & GH & mGGA & 10\% HF & 457 & \citenum{Staroverov2003_12129} \\
      revTPSSh$^{b}$ & GH & mGGA & 10\% HF & 458 & \citenum{Perdew2009_026403, Perdew2009_026403_err, Staroverov2003_12129} \\
      M06 & GH & mGGA & 27\% HF & 449+235 & \citenum{Zhao2008_215} \\
      revM06$^b$ & GH & mGGA & 40.4\% HF & 305+306 & \citenum{Wang2018_10257} \\
      M06-2X & GH & mGGA & 54\% HF & 450+236 & \citenum{Zhao2008_215} \\
      M08-HX & GH & mGGA & 52.2\% HF & 295+78 & \citenum{Zhao2008_1849} \\
      M08-SO & GH & mGGA & 56.8\% HF & 296+77 & \citenum{Zhao2008_1849} \\
      MN15 & GH & mGGA & 44\% HF & 268+269 & \citenum{Yu2016_5032} \\
      M11 & RS & mGGA & 42.8\% SR, 100\% LR & 297+76 & \citenum{Peverati2011_2810} \\
      revM11$^b$ & RS & mGGA & 22.5\% SR, 100\% LR & 304+172 & \citenum{Verma2019_2966} \\
      MN12-SX & RS & mGGA & 25\% SR, 0\% LR & 248+73 & \citenum{Peverati2012_16187} \\
      $\omega$B97M-V & RS & mGGA & 15\% SR, 100\% LR & 531 & \citenum{Mardirossian2016_214110} \\
      \hline
    \end{tabular}
    \begin{tablenotes}
    \item [$^a$] Two numbers indicate the exchange and the correlation functional respectively. A single number indicates an exchange-correlation functional.
    \item [$^b$] Revised version
    \item [$^c$] Regularized version
    \end{tablenotes}
  \end{threeparttable}
  \label{tab:mggas}
\end{table*}

Electronic structure calculations were performed with Hartree--Fock
(HF) and the functionals listed in \tabref{ldas-ggas, mggas} using
\Turbomole{} 7.5\cite{Balasubramani2020_JCP_184107}. Several rungs of
Jacob's ladder were considered when choosing the functionals listed in
\tabref{ldas-ggas, mggas}: local density approximations (LDA),
generalized gradient approximations (GGAs), and meta-GGAs (mGGAs).
Several kinds of functionals are also included: (pure) density
functional approximations, global hybrid (GH) functionals with a
constant amount of HF exchange, as well as range-separated (RS)
hybrids with a given amount of HF exchange in the short range (SR) and
the long range (LR).  As can be seen in \tabref{ldas-ggas, mggas}, the
evaluated functionals consist of one pure LDA, 8 pure GGAs, 8 global
hybrid GGAs, 10 range-separated hybrid GGAs, 12 mGGAs, 8 global hybrid
mGGAs, and 4 range-separated mGGAs, in addition to HF.

The Dunning aug-cc-pCVQZ basis set\cite{Dunning1989_JCP_1007,
  Kendall1992_JCP_6796, Woon1993_JCP_1358, Woon1995_JCP_4572,
  Peterson2002_JCP_10548} (with aug-cc-pVQZ on the hydrogen atoms) and
benchmark quality integration grids were employed in all calculations.
Universal auxiliary basis sets\cite{Weigend2006_PCCP_65} were used
with the resolution-of-the-identity approximation for the Coulomb
interaction in all \Turbomole{} calculations. All density functionals
were evaluated in \Turbomole{} with \Libxc{},\cite{Lehtola2018_S_1}
except the calculations with the recently published CAMh-B3LYP
functional for which \XCFun{} was used.\cite{Ekstroem2010_JCTC_1971}
Magnetizabilities were subsequently evaluated with \Gimic{} by
numerical integration of \eqref{Biot-Savart}. The data necessary for
evaluating the CDT in \Gimic{} were obtained from \Turbomole{}
calculations of nuclear magnetic resonance (NMR) shielding constants
employing GIAOs.\cite{Ditchfield1974_MP_789, Wolinski1990_JACS_8251,
  Kollwitz1998_JCP_8295, Balasubramani2020_JCP_184107,
  Reiter2018_JCTC_191}

Although response calculations are not possible at the moment in the
presence of the non-local correlation kernel used in $\omega$B97X-V,
B97M-V, and $\omega$B97M-V, we have estimated the importance of the
van der Waals (vdW) effects on the magnetic properties by comparing
magnetizabilities obtained with orbitals optimized with and without
the vdW term in the case of \ce{SO2}. The magnetizability obtained
with the vdW optimized orbitals differed by only \jouletesla{0.4}
(0.14\%) from that obtained from a calculation where the vdW term was
omitted in the orbital optimization. Thus, the vdW term appears to
have very little influence on magnetizabilities, as is already
well-known in the literature for other
properties.\cite{Najibi2018_JCTC_5725} The vdW term was therefore not
included in the calculations using the $\omega$B97X-V, B97M-V, and
$\omega$B97M-V functionals in this study.

The accuracy of the numerical integration in \Gimic{} was assessed by
comparing the \Turbomole{}/\Gimic{} magnetizability data to analytical
values from \PySCF{},\cite{Sun2020_JCP_24109} in which
\Libxc{}\cite{Lehtola2018_S_1} was also used to evaluate the density
functionals. Since \PySCF{} does not currently support magnetizability
calculations with mGGA functionals or range-separated functionals,
further calculations were undertaken with \Gaussian{}
16.\cite{Frisch2016__}. The analytical magnetizabilities from \PySCF{}
and \Gaussian{} were found to be in perfect agreement for the studied
LDA and GGA functionals available in both codes (LDA, BP86, PBE, PBE0,
BLYP, B3LYP and BHLYP). Comparison of the data from \PySCF{} to the
\Gimic{} data revealed the numerically integrated magnetizabilities to
be accurate, as the magnetizabilities agreed within \jouletesla{0.5}
for all molecules using the B3LYP, B97-2, B97-3, BLYP, BP86, KT1, KT2,
LDA, PBE, and PBE0 functionals; the small discrepancy may arise from
use of the resolution-of-identity
approximation\cite{Vahtras1993_CPL_514} in \Turbomole{} or from the
numerical integration of the magnetizability density. A comparison of
the raw data for BP86 and B3LYP is given in the SI.

The magnetizabilities calculated with \Gaussian{} and \Turbomole{}
using the meta-GGA functionals were found to differ.  The
discrepancies between the magnetizabilities obtained with the two
programs are due to the use of different approaches to handle the
gauge invariance of the kinetic energy density in meta-GGAs, which are
described in \citerefs{Maximoff2004_CPL_408} and
\citenum{Bates2012_JCP_164105} for \Gaussian{} and \Turbomole{},
respectively. We found the \Turbomole{} data to be significantly
closer to the CCSD(T) reference values.

Finally, since we found the implementation of the KT3 functional in
\Libxc{} version 5.0.0 used by \Turbomole{} to be flawed, the KT3
results in this study are based on calculations with \PySCF{} with a
corrected version of \Libxc{}.

\section{Results}
\label{sec:results}

\subsection{Functional benchmark \label{sec:benchmark}}

\begin{table*}
  \caption{The mean absolute errors (MAEs), mean errors (MEs), and
    standard deviations (STDs) for the magnetizabilities of the 27
    studied molecules in units of $10^{-30}\, {\rm J}/{\rm T}^2$ from
    the CCSD(T) reference with the studied functionals. The
    functionals are ordered in increasing MAE.}
  \label{tab:errors}
\begin{tabular}{llrrr|llrrr}
\hline
Rank & Functional & MAE & ME & STD & Rank & Functional & MAE & ME & STD \tabularnewline
\hline
\hline
 1 &       BHandHLYP &  3.11 &  2.15 &  4.65 & 27 &        revTPSSh &  7.14 &  7.05 &  5.94\tabularnewline
 2 &      CAM-QTP-00 &  3.22 &  0.88 &  4.67 & 28 &           TPSSh &  7.20 &  7.07 &  6.02\tabularnewline
 3 &  $\omega$B97X-V &  3.22 &  2.51 &  4.36 & 29 &           B97-2 &  7.24 &  7.07 &  6.40\tabularnewline
 4 &      CAM-QTP-01 &  3.23 &  0.59 &  4.49 & 30 &          M08-HX &  7.34 &  5.17 & 10.27\tabularnewline
 5 &      CAM-QTP-02 &  3.28 & -0.23 &  4.36 & 31 &            BLYP &  7.91 &  5.69 &  8.75\tabularnewline
 6 &     $\omega$B97 &  3.54 &  2.44 &  4.75 & 32 &          N12-SX &  8.04 &  7.89 &  7.48\tabularnewline
 7 &  $\omega$B97M-V &  3.61 &  0.41 &  4.75 & 33 &         revTPSS &  8.20 &  7.86 &  6.68\tabularnewline
 8 &       CAM-B3LYP &  3.73 &  2.38 &  4.86 & 34 &            TPSS &  8.22 &  7.85 &  6.85\tabularnewline
 9 &         MN12-SX &  3.80 &  0.22 &  5.34 & 35 &          revM11 &  8.23 &  6.83 & 10.03\tabularnewline
10 &      CAMh-B3LYP &  4.23 &  3.22 &  5.17 & 36 &            TASK &  8.27 &  7.31 &  7.43\tabularnewline
11 &    $\omega$B97X &  4.25 &  3.71 &  5.22 & 37 &            BP86 &  8.59 &  7.30 &  8.75\tabularnewline
12 &          QTP-17 &  4.58 &  3.77 &  5.45 & 38 &           M11-L &  8.92 &  5.20 &  9.26\tabularnewline
13 &           BHLYP &  4.73 &  0.10 &  6.47 & 39 &          revM06 &  8.94 &  8.67 & 10.27\tabularnewline
14 &          B97M-V &  5.19 &  4.13 &  5.58 & 40 &             PBE &  9.13 &  7.07 &  9.42\tabularnewline
15 &        revB3LYP &  5.45 &  4.34 &  6.13 & 41 &             KT3 &  9.19 &  8.38 &  8.08\tabularnewline
16 &           B3LYP &  5.47 &  4.72 &  5.97 & 42 &             LDA &  9.55 &  5.37 & 11.36\tabularnewline
17 &          MN12-L &  5.79 & -2.03 &  8.02 & 43 &        CHACHIYO &  9.76 &  9.17 &  8.88\tabularnewline
18 &             KT1 &  5.87 &  1.15 &  7.11 & 44 &             M11 &  9.93 &  7.61 & 13.77\tabularnewline
19 &           rSCAN &  5.91 &  5.00 &  6.06 & 45 &          M06-2X & 10.15 &  9.01 & 13.12\tabularnewline
20 &            PBE0 &  5.96 &  5.56 &  6.81 & 46 &             MVS & 10.35 &  9.92 &  9.20\tabularnewline
21 &  $\omega$B97X-D &  6.22 &  5.89 &  6.35 & 47 &          M08-SO & 10.40 &  8.09 & 14.34\tabularnewline
22 &            SCAN &  6.30 &  5.89 &  5.96 & 48 &             N12 & 10.89 & 10.01 &  9.58\tabularnewline
23 &             KT2 &  6.42 &  5.58 &  7.21 & 49 &            MN15 & 11.45 & 10.45 & 12.82\tabularnewline
24 &          MN15-L &  6.57 & -5.27 &  6.94 & 50 &           M06-L & 12.49 & 12.45 &  9.42\tabularnewline
25 &           B97-3 &  6.61 &  6.61 &  6.26 & 51 &             M06 & 13.34 & 13.11 & 13.16\tabularnewline
26 &        revM06-L &  7.00 &  6.23 &  5.98 & 52 &              HF & 18.40 &  7.48 & 61.81\tabularnewline
\hline
\end{tabular}

\end{table*}

\begin{figure*}
  \subfigure[]{
    \includegraphics[width=0.31\textwidth]{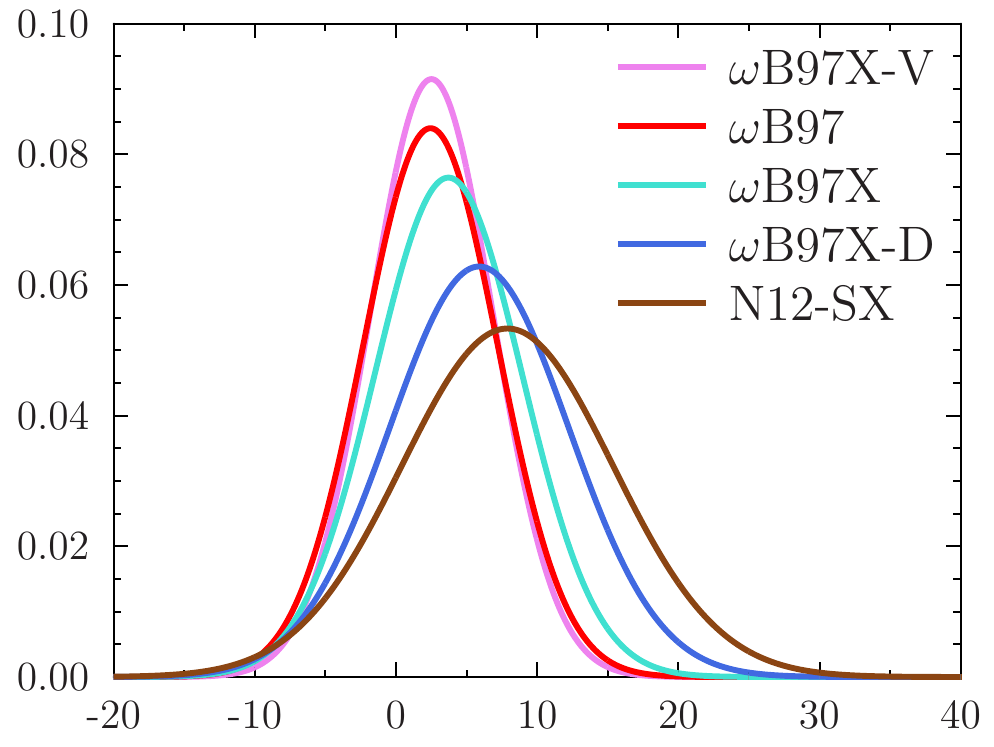}
    \label{fig:figure1a}}
  \subfigure[]{
    \includegraphics[width=0.31\textwidth]{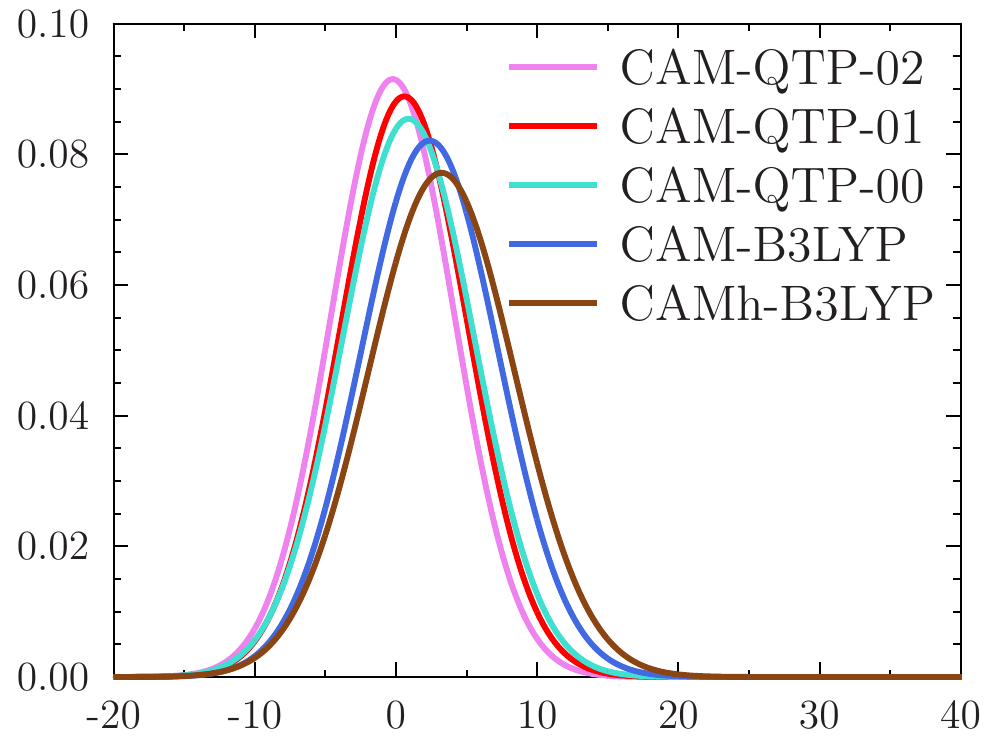}
    \label{fig:figure1b}}
  \subfigure[]{
    \includegraphics[width=0.31\textwidth]{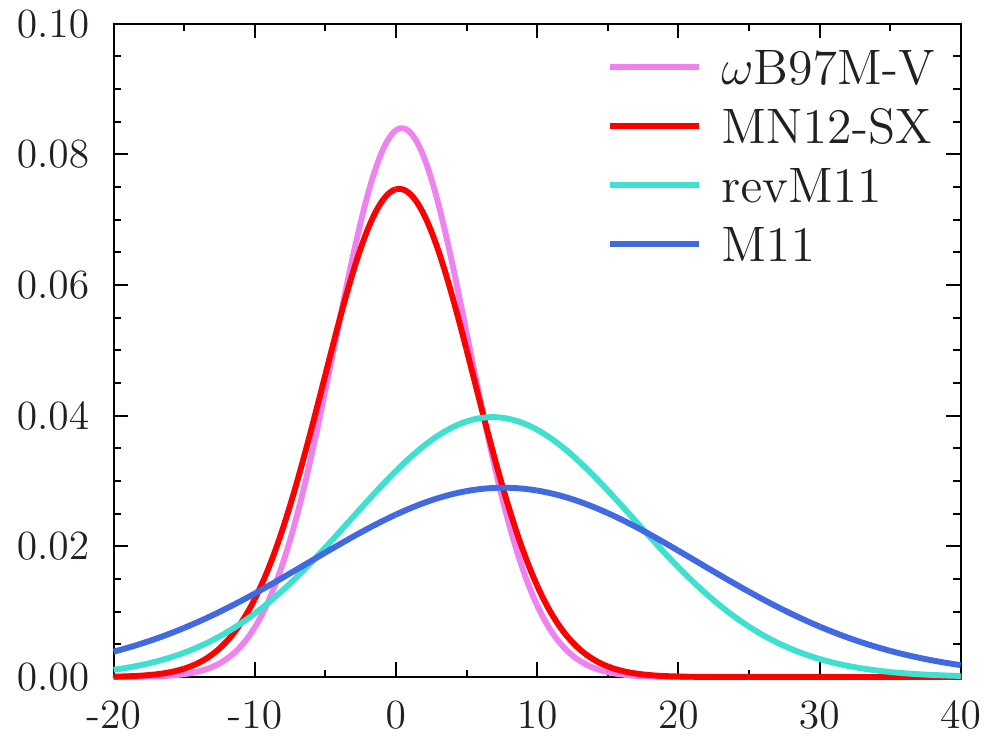}
    \label{fig:figure1c}}\\
  \subfigure[]{
    \includegraphics[width=0.31\textwidth]{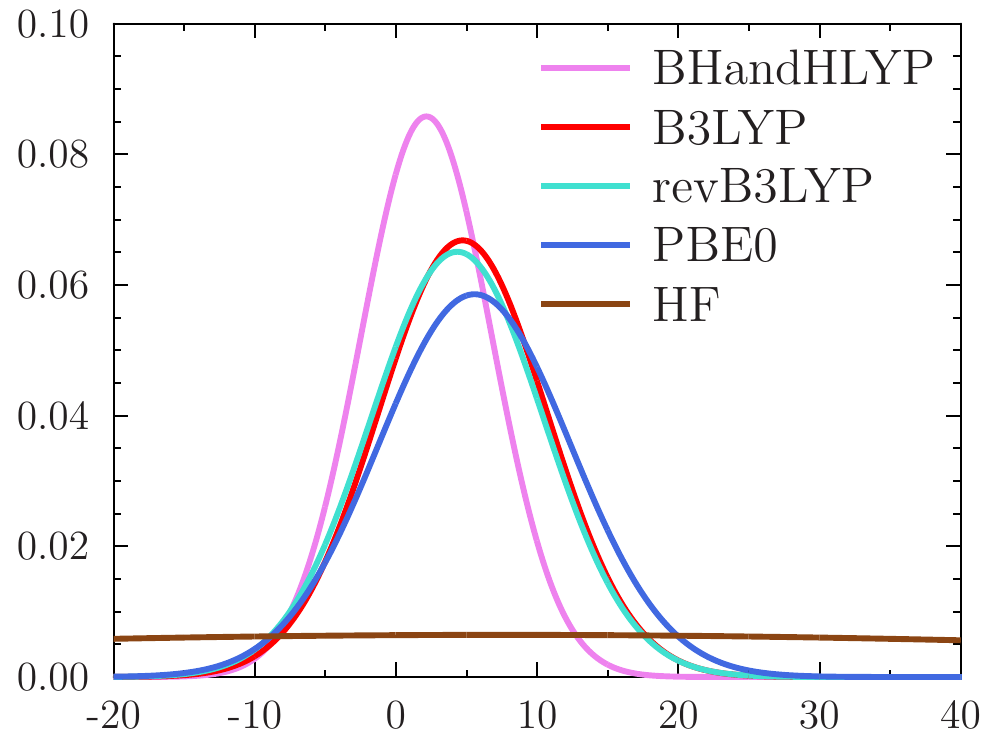}
    \label{fig:figure1d}}
  \subfigure[]{
    \includegraphics[width=0.31\textwidth]{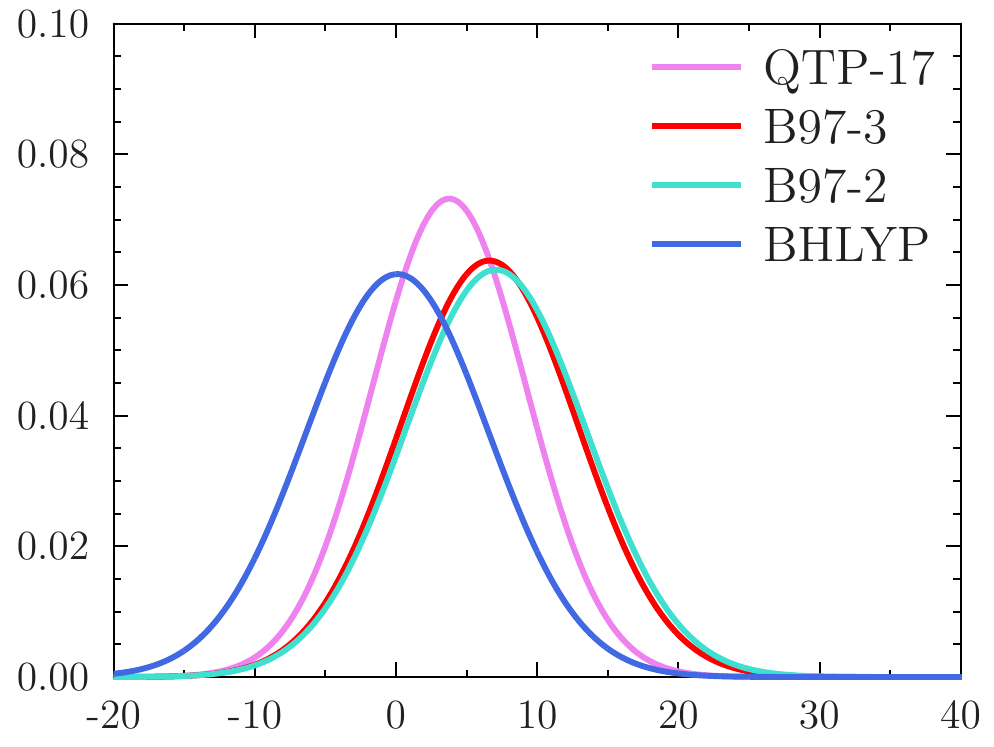}
    \label{fig:figure1e}
  }
  \subfigure[]{
    \includegraphics[width=0.31\textwidth]{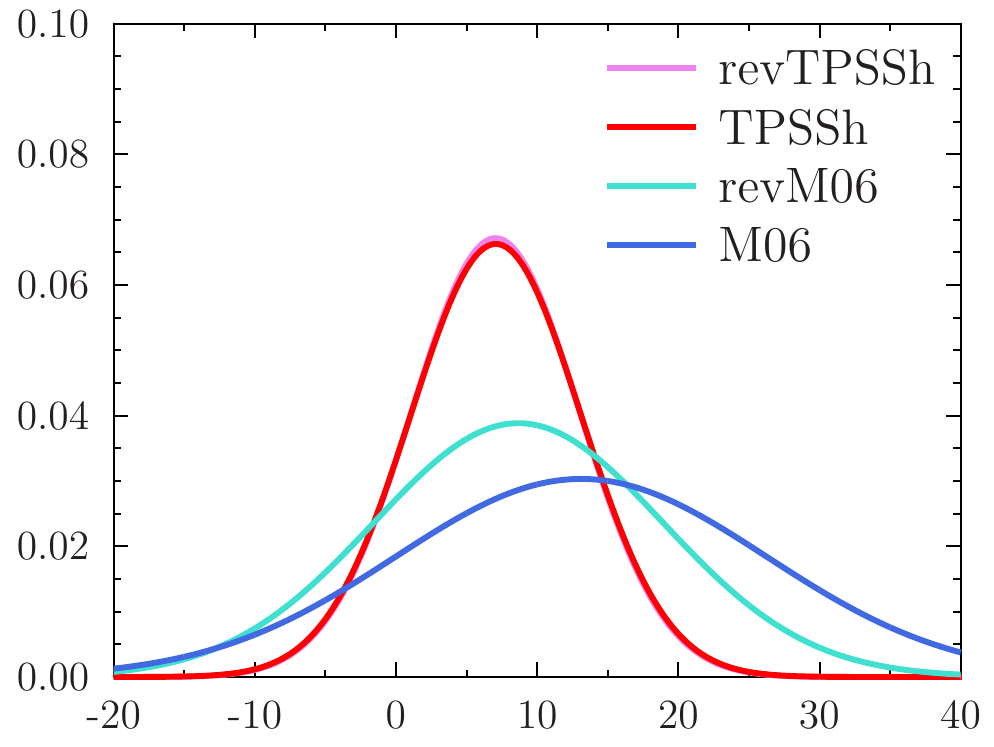}
    \label{fig:figure1f}
  }\\
  \subfigure[]{
    \includegraphics[width=0.31\textwidth]{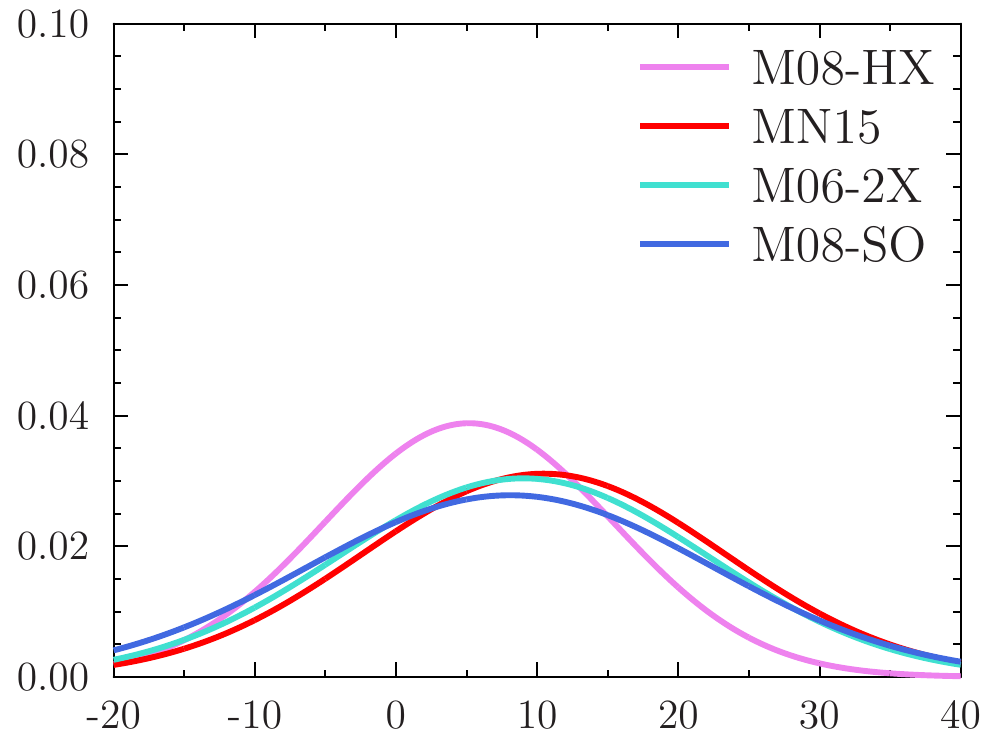}
    \label{fig:figure1g}
  }
  \subfigure[]{
    \includegraphics[width=0.31\textwidth]{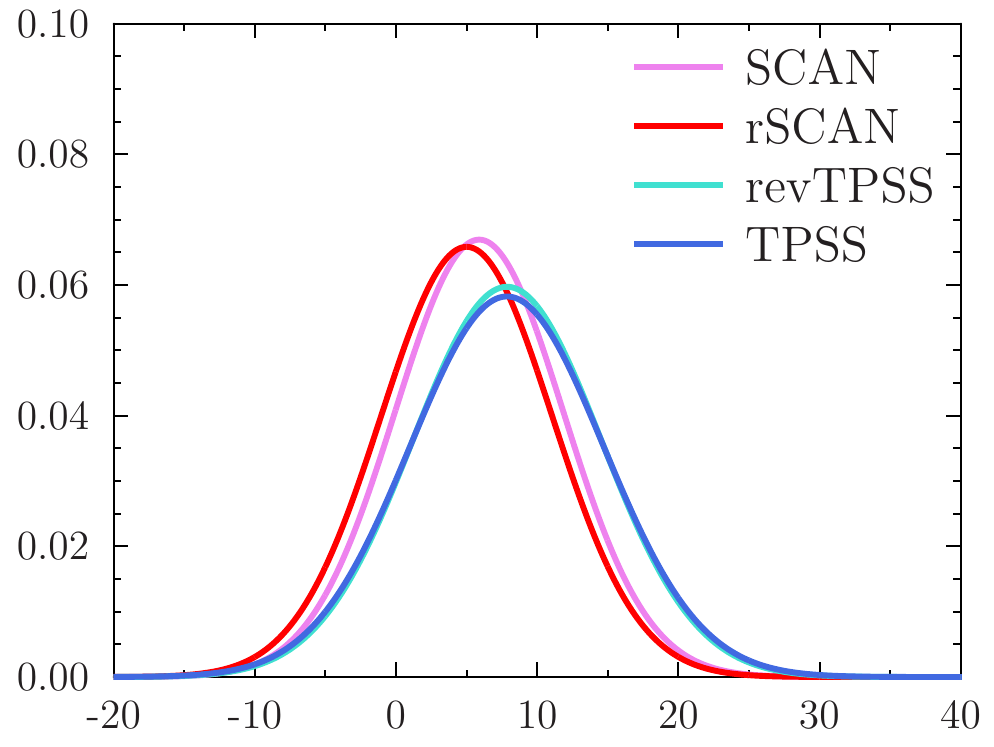}
    \label{fig:figure1h}
  }
  \subfigure[]{
    \includegraphics[width=0.31\textwidth]{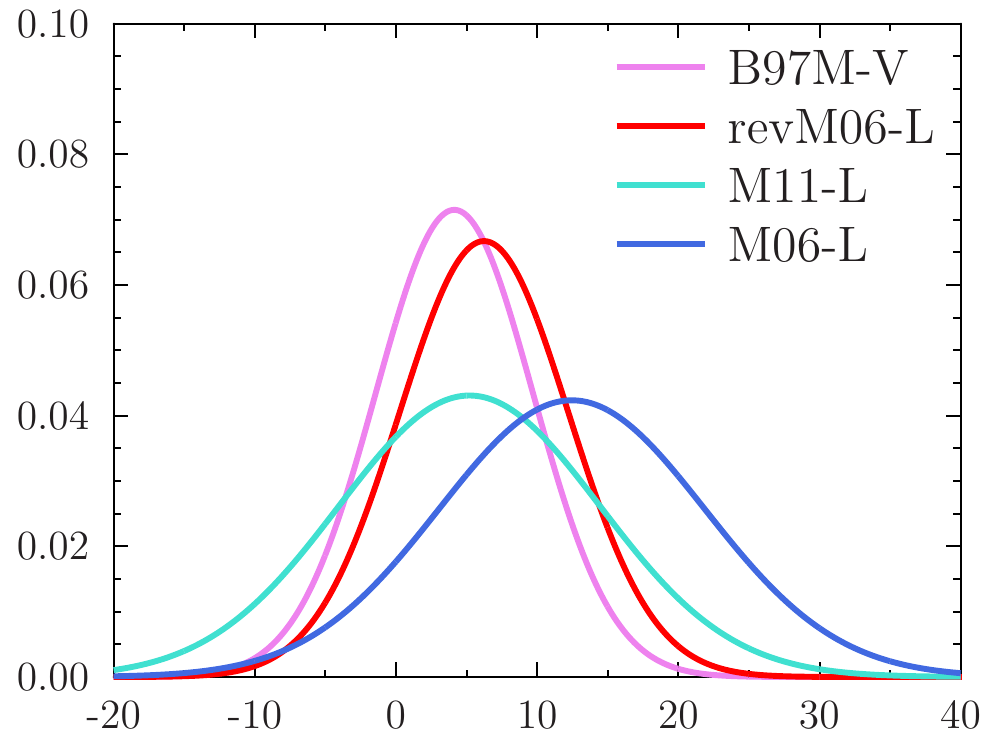}
    \label{fig:figure1i}
  } \\
  \subfigure[]{
    \includegraphics[width=0.31\textwidth]{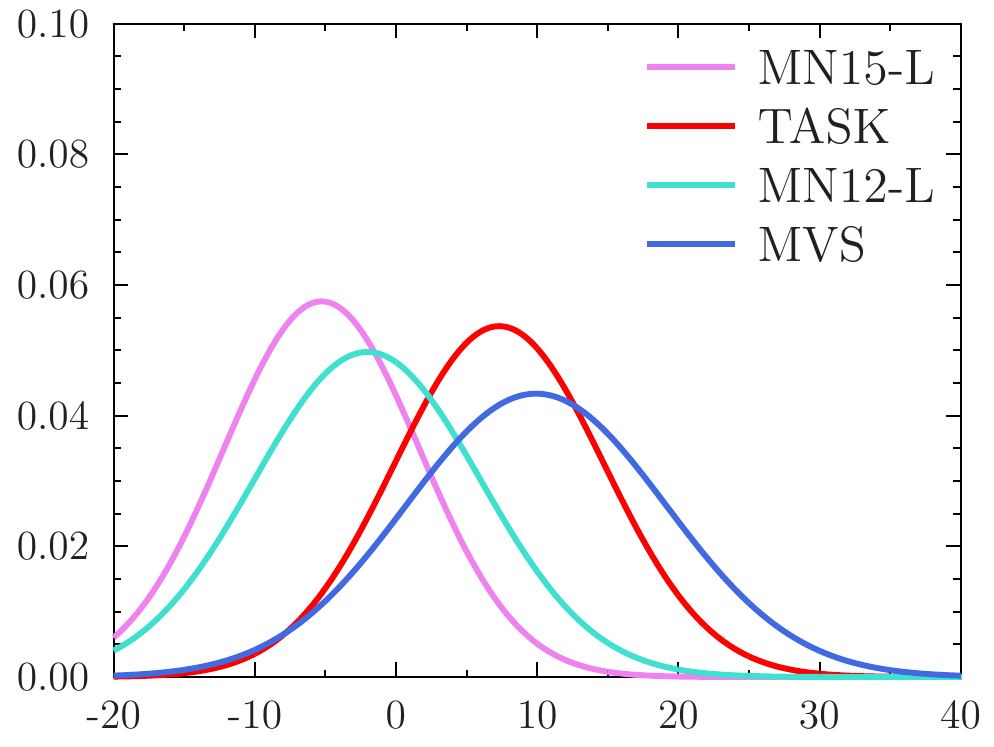}
    \label{fig:figure1j}
  }
  \subfigure[]{
    \includegraphics[width=0.31\textwidth]{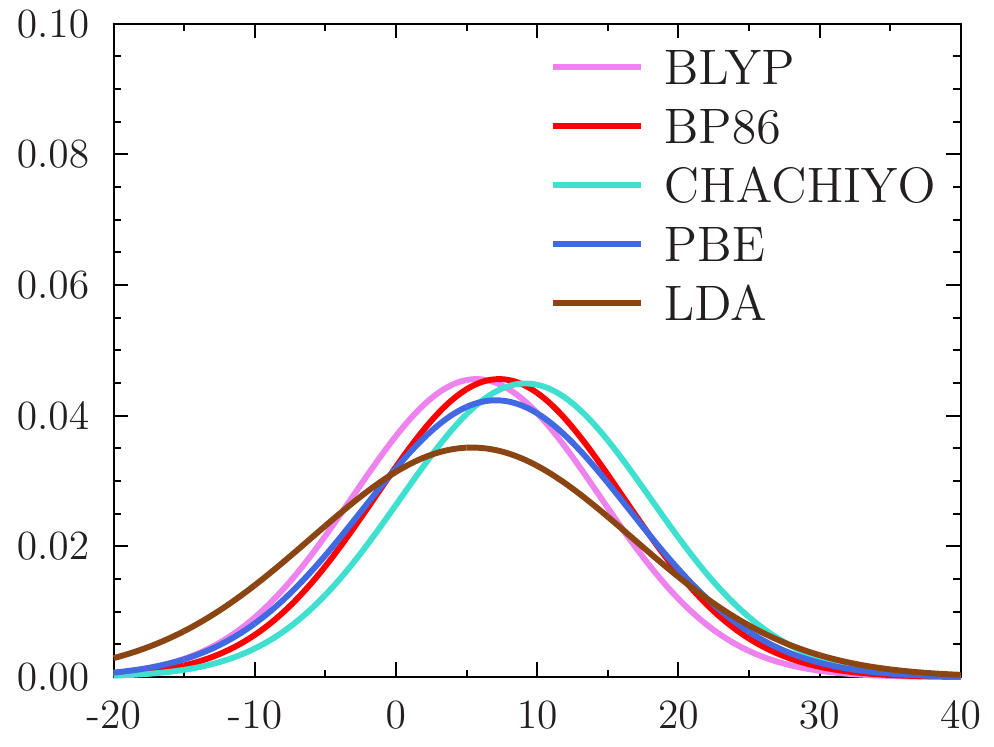}
    \label{fig:figure1k}
  }
  \subfigure[]{
    \includegraphics[width=0.31\textwidth]{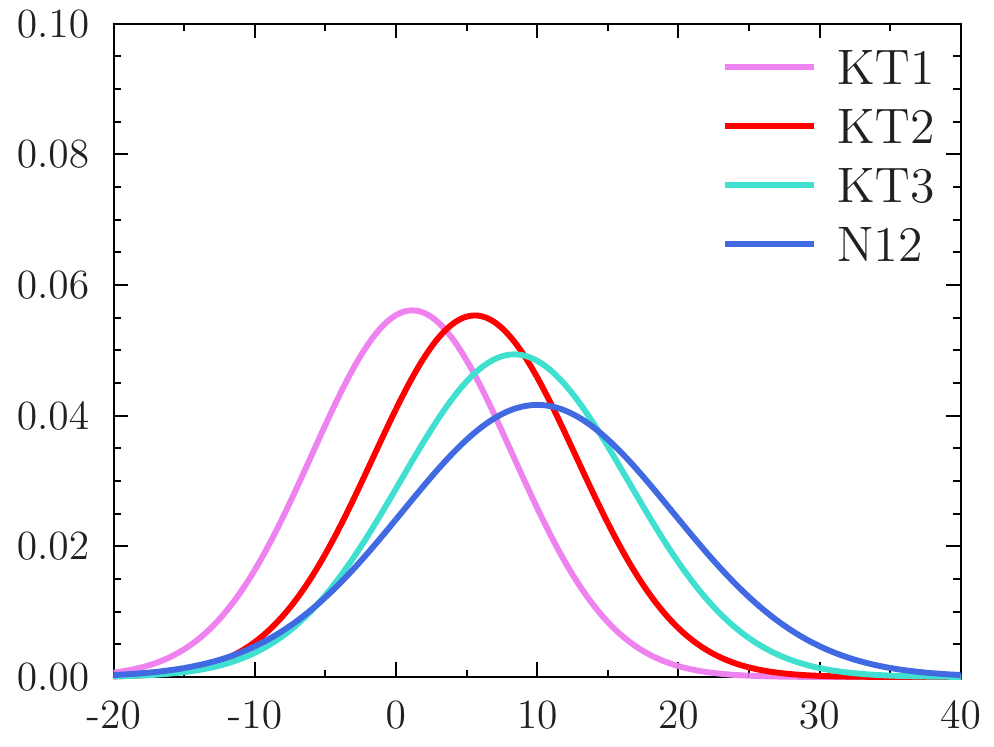}
    \label{fig:figure1l}
  }

\caption{Normal distributions (ND) representing the errors in the
  magnetizabilities for the 27 benchmark reproduced by the studied
  functionals, obtained by plotting the data presented in
  \tabref{errors}. The curves are ordered in each figure by increasing
  standard deviation. The NDs of RS functionals are shown in
  \figref{figure1a, figure1b, figure1c}. The NDs of the GH functionals
  are shown in \figref{figure1d, figure1e, figure1f, figure1g}. The
  NDs of the mGGA functionals are shown in \figref{figure1h, figure1i,
    figure1j}. The NDs of the LDA and GGA functionals are shown in
  \figref{figure1k, figure1l}.  }
  \label{fig:normal}
\end{figure*}

The deviations of the DFT magnetizabilities from the CCSD(T) reference values
of \citeref{Lutnes2009_JCP_144104} are visualized as ideal normal distributions (NDs) in
\figref{normal}. The visualization shows the idealized distribution of the
error in the magnetizability for each functional, based on the computed mean
errors (ME) and standard deviation of the error (STD) given in \tabref{errors}.
The raw data on the magnetizabilities and the differences from the CCSD(T)
reference are available in the SI. Although the error distributions in
\figref{normal} are instructive, we will employ mean absolute errors (MAEs) in
order to rank the functionals studied in this work in a simple, unambiguous
fashion. The MAEs are also given in \tabref{errors}.

Examination of the data in \tabref{errors} shows that range-separated
(RS) functionals generally yield accurate magnetizabilities. Judged by
the mean absolute error, the best performance is obtained with the
BHandHLYP GH functional.  BHandHLYP is followed by 10 RS
functionals, which have much sharper distributions than the
rest of the studied functionals. The best performing
RS functionals are three of the six Berkeley RS functionals
($\omega$B97X-V, $\omega$B97, $\omega$B97M-V) and the three RS
functionals from the University of Florida's Quantum Theory Project
(QTP) CAM-QTP-00, CAM-QTP-01, and CAM-QTP-02. Five of these
functionals have 100\% long-range (LR) HF exchange, while the
CAM-QTP-00 functional has 91\% LR HF exchange.  The two other RS
Berkeley functionals with 100\% LR exchange are ranked \rankth{11}
($\omega$B97X) and \rankst{21} ($\omega$B97X-D) among the studied
functionals. The NDs of the studied RS GGA functionals are shown in
\figref{figure1a, figure1b}, whereas the NDs of the studied RS mGGA
functionals are shown in \figref{figure1c}.

The CAM-B3LYP (65\% LR HF exchange) and CAMh-B3LYP (50\% LR HF
exchange) functionals are among the top ten functionals (ranked
\rankth{8} and 10$^\mathrm{th}$, respectively). CAM-B3LYP was designed
for the accurate description of charge transfer excitations in a
dipeptide model,\cite{Yanai2004_51} while CAMh-B3LYP functional is
aimed at excitation energies of biochromophores.\cite{Shao2020_587}

The best Minnesota functional, MN12-SX, is ranked \rankth{9}. MN12-SX
is a highly parameterized functional with 58 parameters that is known
to require the use of extremely accurate integration
grids.\cite{Mardirossian2016_JCTC_4303} Furthermore, since MN12-SX is
a RS functional with HF exchange only in the short range (SR), it may
have problems modeling magnetic properties of antiaromatic molecules
sustaining strong ring currents in the paratropic (nonclassical)
direction.\cite{Valiev2017_CC_9866, Valiev2018_JPCA_4756,
  Valiev2020_JPCC_21027} We illustrate this with calculations on the
strongly antiaromatic tetraoxa-isophlorin molecule in the Supporting
Information: MN12-SX yields a magnetizability that is four times
larger than the LMP2 [local second-order M\o{}ller--Plesset
  perturbation theory] reference value, while the magnetizabilities
from BHandHLYP and CAM-B3LYP are in good agreement with LMP2. The
N12-SX functional ranked \ranknd{32} is also a RS functional with 0\%
LR exchange.  The RS Minnesota functionals with 100\% LR HF exchange
(M11 and revM11) have large MAEs of \jouletesla{9.93} and
\jouletesla{8.87} and are ranked \rankth{44} and \rankth{35},
respectively.

The best global hybrid (GH) functional is BHandHLYP, which is ranked
\rankst{1} among all functionals of this study, as was already
mentioned above. Among GHs, BHandHLYP is followed by QTP-17, which is
ranked \rankth{12}. Old and established GH functionals like BHLYP
a.k.a. BHandH, B3LYP, and PBE0 perform almost as well as QTP-17 and are
ranked \rankth{13}, \rankth{16}, and \rankth{20},
respectively. The performance of revB3LYP is practically the same as
for B3LYP; the same holds for revTPSSh and TPSSh. The other
established GH functionals like B97-2, B97-3, TPSSh and newer ones
like revTPSSh and M08-HX are found in the beginning of the second half
of the ranking list, whereas M08-SO, M06, revM06, M06-2X, MN15, and
M06 are ranked between \rankth{39} and \rankst{51}. The NDs
of the GH functionals are compared in \figref{figure1d, figure1e,
  figure1f, figure1g}.

B97M-V, at the \rankth{14} place, is the best pure mGGA
functional. The rSCAN and SCAN functionals are ranked \rankth{19} and
\rankth{22}, respectively, whereas revTPSS and TPSS appear at
positions 33 and 34, respectively. The pure mGGA functionals of the
Minnesota series are ranked \rankth{17} (MN12-L), \rankth{24}
(MN15-L), \rankth{26} (revM06-L), and \rankth{50} (M06-L). The
performance of the Minnesota pure mGGA functionals, excluding M06-L,
is about the same as that of TASK and the other mGGA functionals. The
magnetizabilities calculated with the revised M06-L (revM06-L)
functional are more accurate than those with M06-L. The MVS mGGA
functional is ranked \rankth{46}.  The NDs for the mGGA functionals
are shown in \figref{figure1h, figure1i, figure1j}.

The magnetizabilities calculated with several of the Minnesota functionals are
inaccurate. Seven of the eight worst performing functionals (M11, M06-2X, MVS,
M08-SO, N12, MN15, M06-L, M06) in \tabref{errors} are Minnesota functionals.
Five other Minnesota functionals are also ranked in the lower half, placing
\rankth{30} (M08-HX), \rankth{32} (N12-SX), \rankth{35} (revM11), \rankth{38}
(M11-L), and \rankth{39} (revM06).

The KT1 and KT2 functionals are the best GGA functionals, ranking
\rankth{18} and \rankrd{23}, respectively; both KT1 and KT2 have been
optimized for NMR shieldings.\cite{Keal2003_3015} The older
commonly-used GGAs \ie{}, BLYP, BP86, and PBE are ranked \rankst{31},
\rankth{37}, and \rankth{40}, respectively, which is only slightly
better than KT3 ranked \rankst{41} and LDA ranked \ranknd{42}. The
CHACHIYO and N12 functionals, which are newer GGAs, are ranked
\rankrd{43} and \rankth{48}, respectively. The NDs of the GGA
functionals and the LDA are shown in \figref{figure1k, figure1l}.

The magnetizabilities calculated at the HF level are significantly
less accurate and have a much larger MAE-STD than those obtained at
the DFT levels, and we cannot recommend the use of HF for
magnetic properties.

\begin{figure*}
  \subfigure[]{\includegraphics[height=38mm]{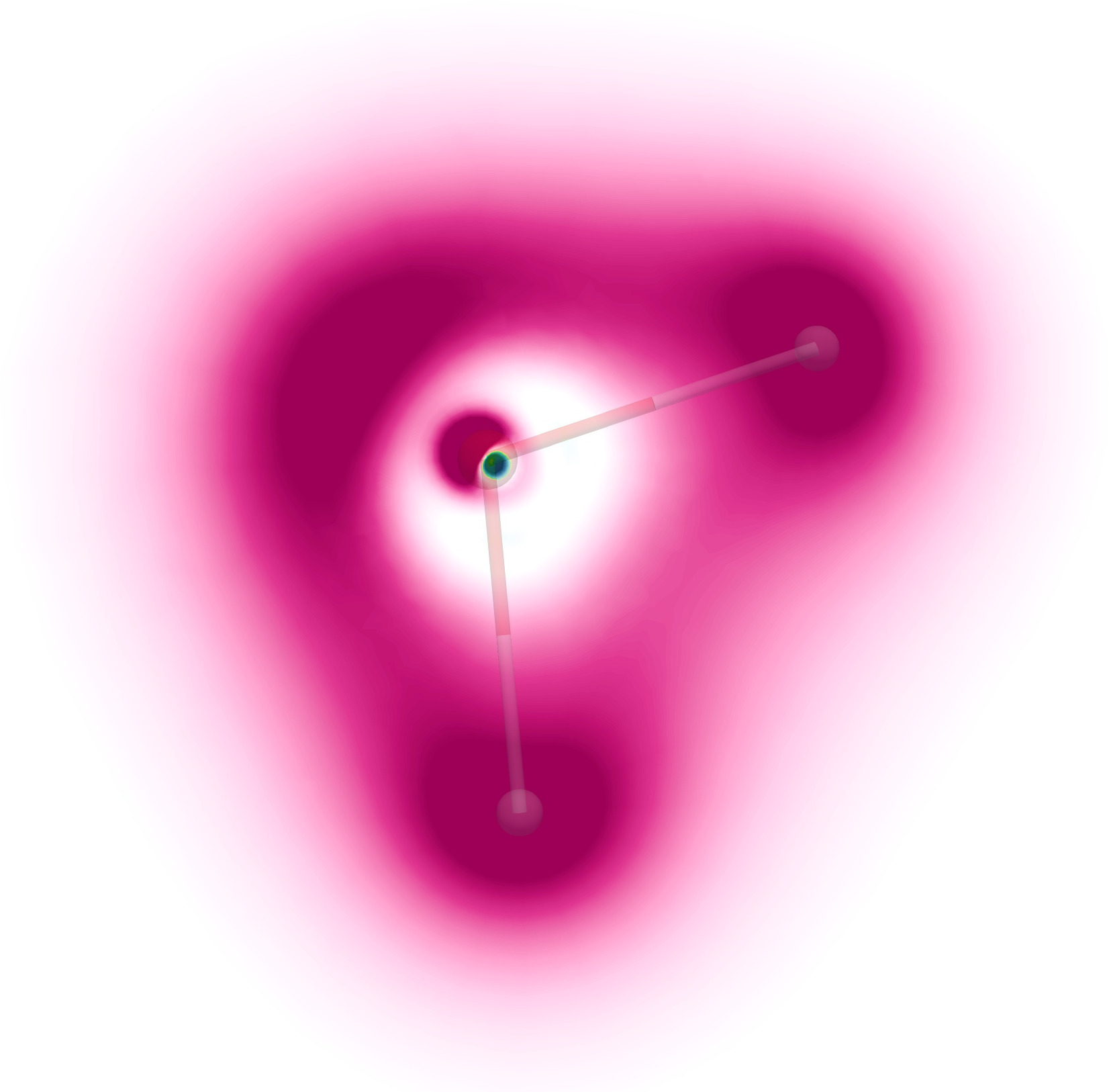} \label{fig:magnetizability-function-h2o}}
  \subfigure[]{\includegraphics[height=38mm]{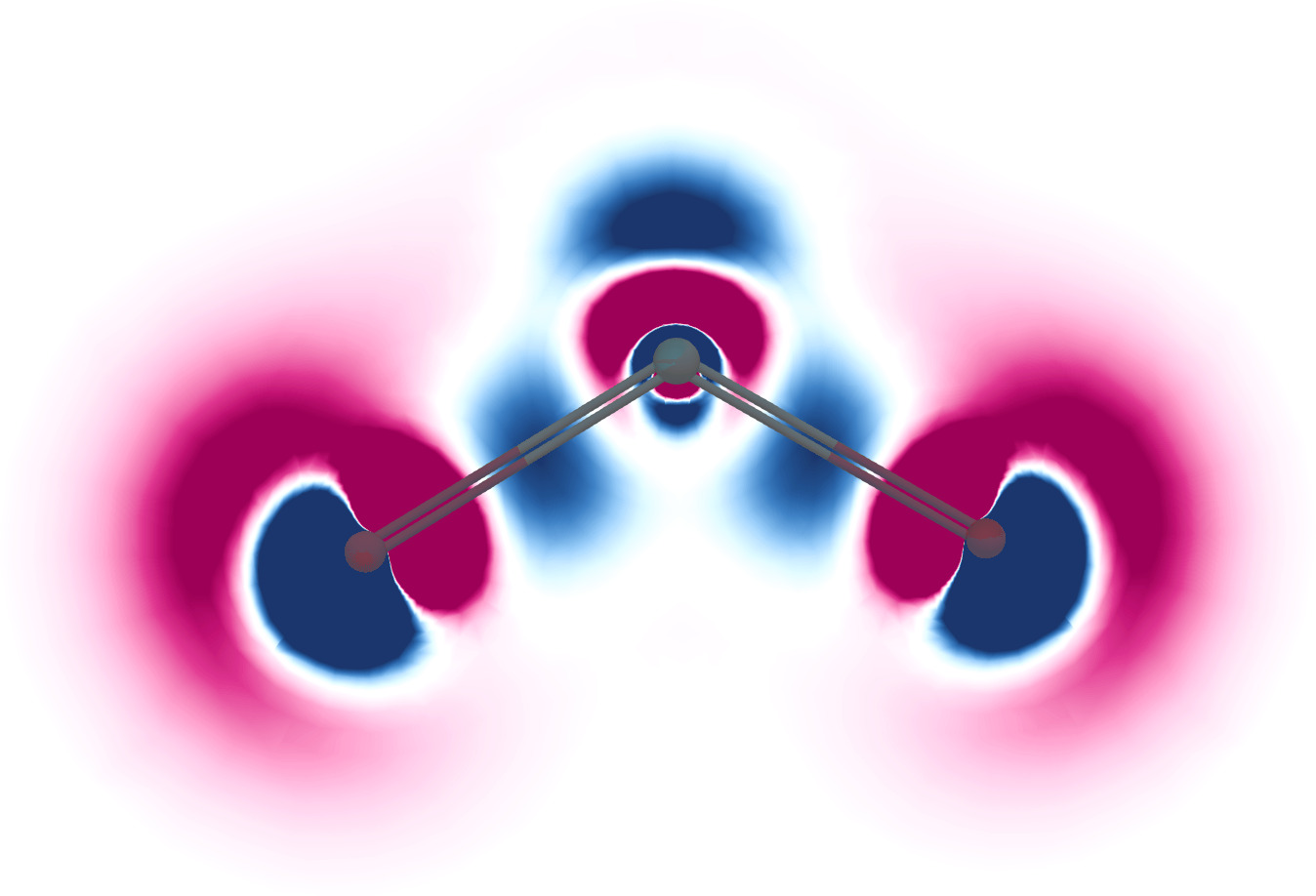} \label{fig:magnetizability-function-so2}}
  \subfigure[]{\includegraphics[height=38mm]{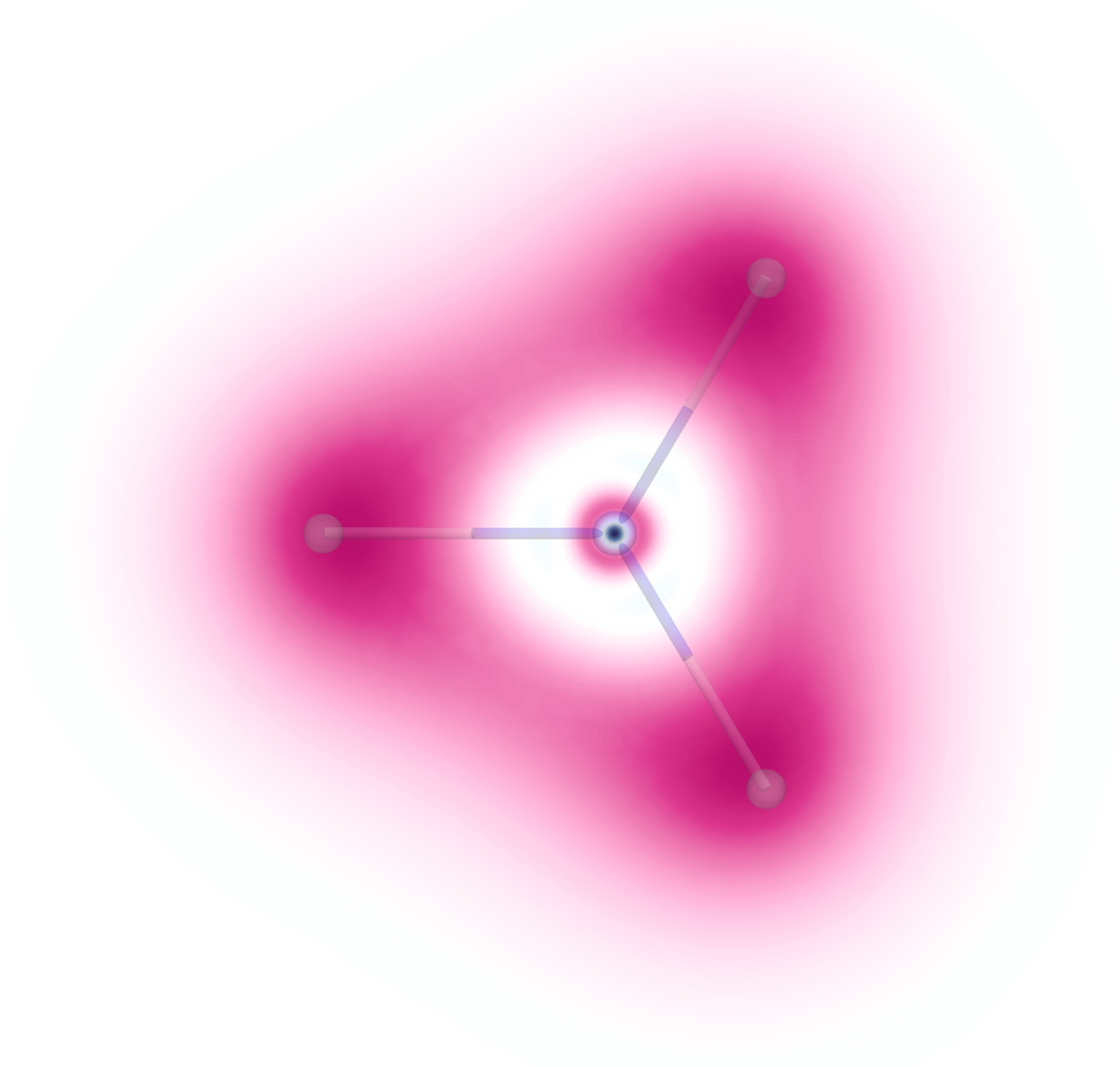} \label{fig:magnetizability-function-nh3}}

\caption{Visualization of the isotropic magnetizability density
$\overline{\rho}^\xi({\bf r})$ (\eqref{isomagdens}) shown in the molecular
plane of \ce{H2O} \ref{fig:magnetizability-function-h2o} and \ce{SO2}
\ref{fig:magnetizability-function-so2} as well as in the plane formed by the
hydrogen atoms of \ce{NH3} \ref{fig:magnetizability-function-nh3}, positioned
\bohr{0.06} away from the \ce{N} atom towards the hydrogen atoms.  Negative
contributions are shown in pink, and positive ones in green. The gauge origin
$\mathbf{R}_O$ is $(0,0,0) \bohr{}$}.
\label{fig:magnetizability-function}
\end{figure*}

\subsection{Magnetizability densities \label{sec:magnetizability}}

\begin{figure*}
  \subfigure[]{\includegraphics[width=0.30\linewidth]{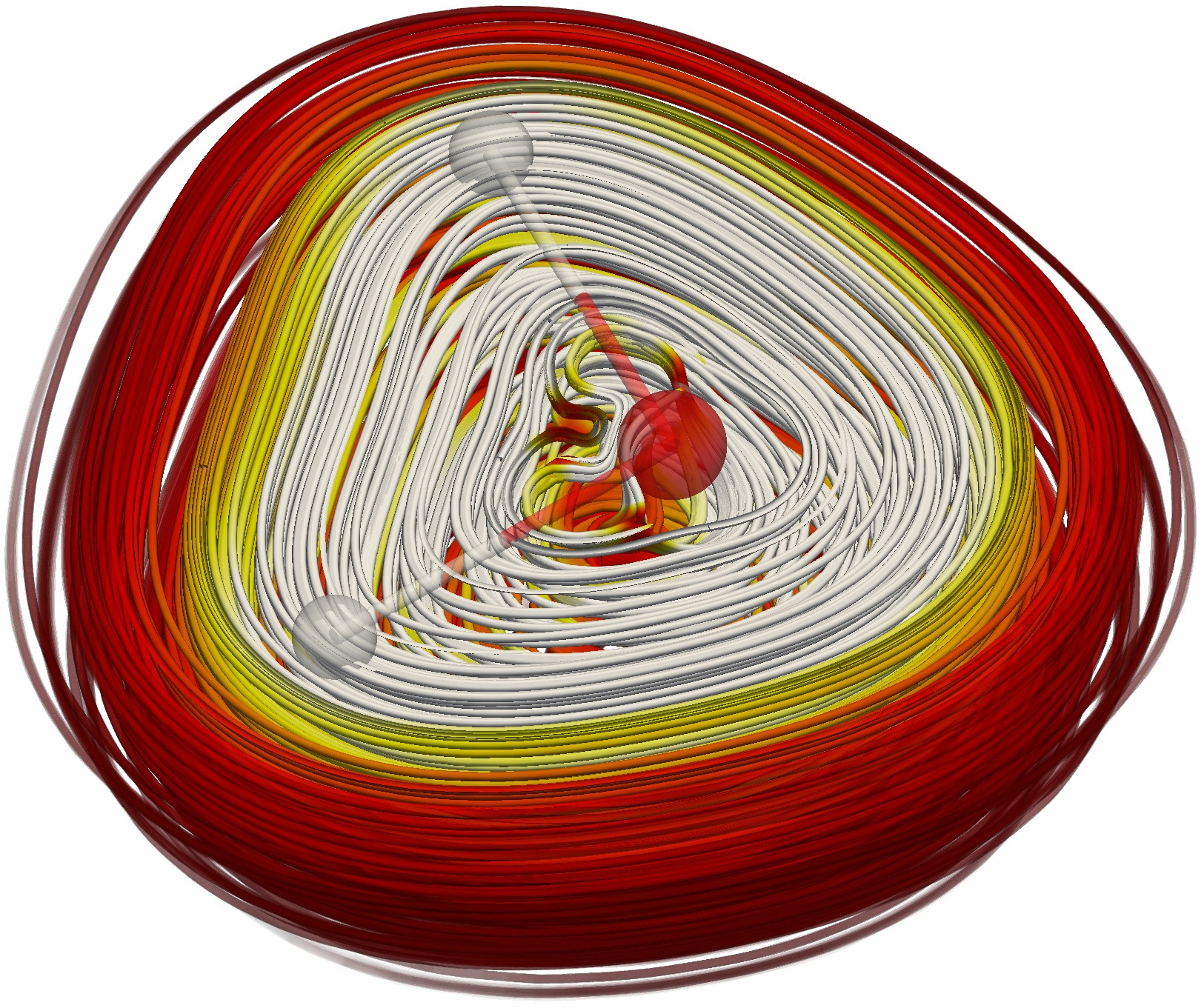} \label{fig:spaghetti-h2o}}
  \vspace{12mm}
  \subfigure[]{\includegraphics[width=0.28\linewidth]{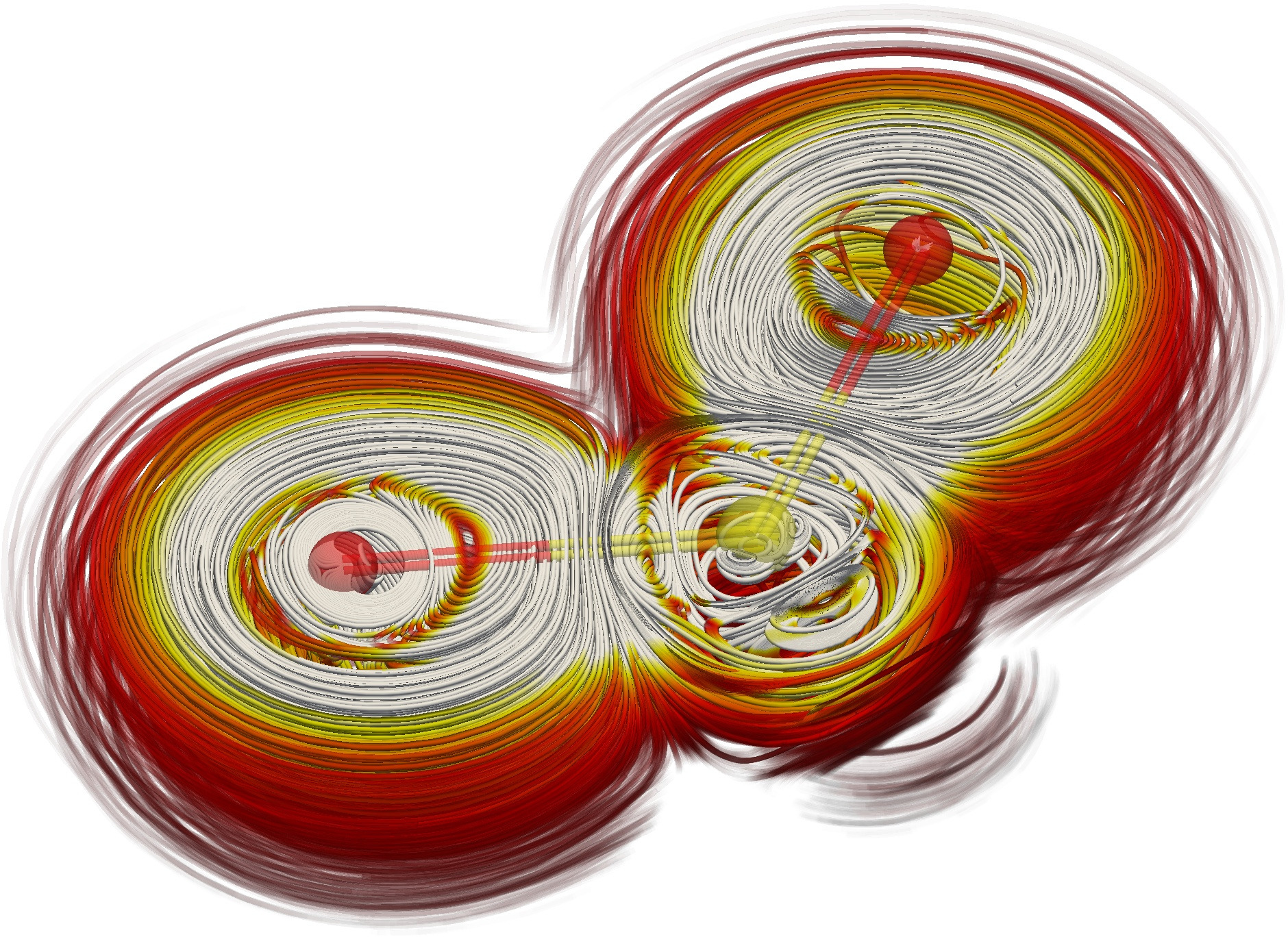} \label{fig:spaghetti-so2}}
  \subfigure[]{\includegraphics[width=0.19\linewidth]{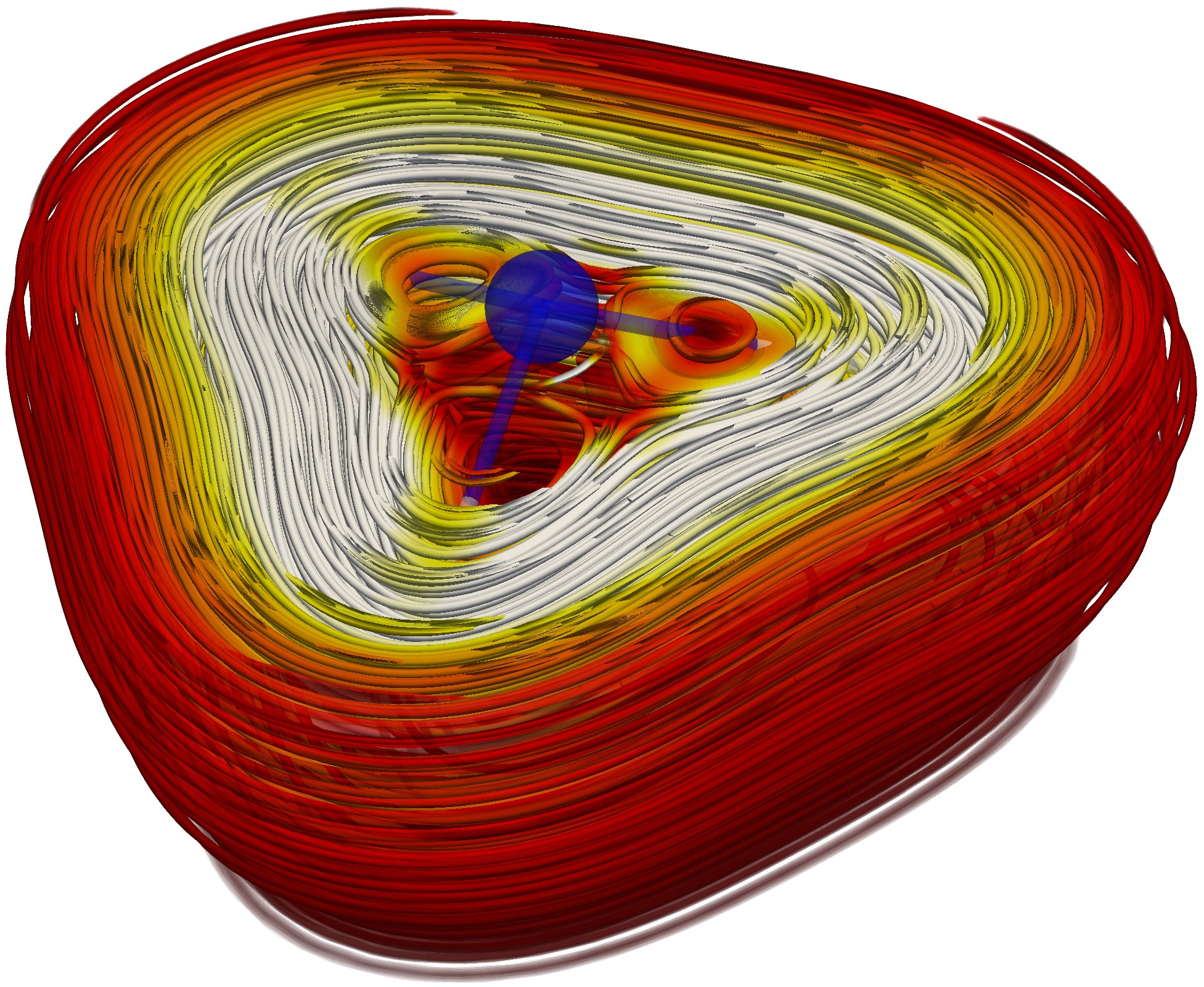} \label{fig:spaghetti-nh3}}
  \vspace{5mm}

\caption{Streamline representation of the CDT (\eqref{MCDSS}) of \ce{H2O}
(\ref{fig:spaghetti-h2o}), \ce{SO2} (\ref{fig:spaghetti-so2}) and \ce{NH3}
(\ref{fig:spaghetti-nh3}). The CDT is calculated with the magnetic field
perpendicular to the molecular plane of \ce{H2O} and \ce{SO2} as well as with
it along the symmetry axis of \ce{NH3}.  The color scale represents the
strength of the CDT in \nATbohr{}.} \label{fig:spaghetti}

\end{figure*}

Spatial contributions to the magnetizability densities, \ie{}, the
integrand in \eqref{Biot-Savart}, are illustrated for \ce{H2O},
\ce{NH3} and \ce{SO2} in \figref{magnetizability-function}, with
\figref{spaghetti} showing the corresponding CDTs. The
magnetizability densities are calculated with the gauge origin of the
external magnetic field $(\mathbf{R}_O)$ at $(x,y,z)=(0,0,0)$. In the
calculations on \ce{H2O} and \ce{SO2}, the magnetic field perturbation
is perpendicular to the molecular plane, while for \ce{NH3} the
perturbation is parallel to the $C_3$ symmetry axis. In the case of
\ce{H2O}, the current-density flux around the whole molecule
(\figref{spaghetti-h2o}) leads to the ring-shaped contribution shown
in \figref{magnetizability-function-h2o}. The magnetic field along the
symmetry axis of \ce{NH3} also results in a current-density flux
around the molecule at the hydrogen atoms (\figref{spaghetti-nh3}),
giving rise to a similar ring-shaped contribution shown in
\figref{magnetizability-function-nh3}.

The isotropic magnetizability density of \ce{SO2} shown in
\figref{magnetizability-function-so2} has positive (green) and
negative (pink) values.  Calculations of the CDT show that the
oxygens sustain a strong diatropic atomic CDT that flows around the
atom, whereas the atomic CDT of the sulfur atom is much weaker
(\figref{spaghetti-so2}).  The $p$-orbital shaped contributions to the
magnetizability density of \ce{SO2} around the oxygens in
\figref{magnetizability-function-so2} originate from the atomic
CDTs. The patterns of the CDT of \ce{H2O} and \ce{SO2} lead to the
different magnetizability densities seen in
\figref{magnetizability-function-h2o, magnetizability-function-so2},
respectively.  The positive magnetizability densities in \ce{H2O} and
\ce{NH3} are extremely localized close to the atomic nuclei, also
because of vortices of the atomic CDT.

The magnetizability density depends on the gauge origin of the vector
potential of the external magnetic field, even though the
magnetizability is independent of the gauge
origin.\cite{Jameson1980_JCP_5684} The magnetizability densities for
\ce{H2O}, \ce{NH3} and \ce{SO2} calculated with the gauge origin at
$\mathbf{R}_O = (1,1,1) \bohr{}$ are shown in the SI.  The
contribution of the choice of the gauge origin to the magnetizability
computed from \eqref{Biot-Savart} vanishes when the CDT fulfills the
charge conservation condition\cite{Sambe1973_JCP_555}
\begin{equation}
  \int {\cal J}_\alpha^{B_\beta}(\mathbf{r}) \mathrm{d}^3 r = 0.
  \label{eq:conservation}
\end{equation}
Calculating the magnetizability for \ce{NH3} with a gauge origin set
to $\mathbf{R}_O = (100,100,100) \bohr{}$ yielded a value that differs
by 0.32\% from the one computed for $\mathbf{R}_O = (0,0,0)$. When the
gauge origin is set to $\mathbf{R}_O = (1,1,1) \bohr{}$, the deviation
is two orders of magnitude smaller, because the change in the
magnetizability depends linearly on the relative position of the gauge
origin. The magnetizabilities of \ce{H2O} and \ce{SO2} also change by
only 0.46\% and 0.03\% when moving the gauge origin from $(0,0,0)
\bohr{}$ to $(100,100,100) \bohr{}$, respectively, showing that that
charge conservation is practically fulfilled in our calculations.  All
other positions than $(0,0,0)$ for the gauge origin lead to a spurious
CDT contribution to the magnetizability density.

The GIAO ansatz modifies the atomic orbitals leading to a magnetic
response of an external magnetic field that is correct to the first
order for the one-center problem.\cite{Lazzeretti2000_PNMRS_1,
  Magyarfalvi2011__} Even though they do not guarantee that the
integral condition for the charge conservation of the CDT is
fulfilled,\cite{Epstein1973_JCP_1592} the basis set convergence is
faster and the leakage of the CDT is much smaller when GIAOs are
used.\cite{Juselius2004_JCP_3952}

\section{Conclusions}
\label{sec:conclusions}

\begin{figure*}
  \includegraphics[width=\textwidth]{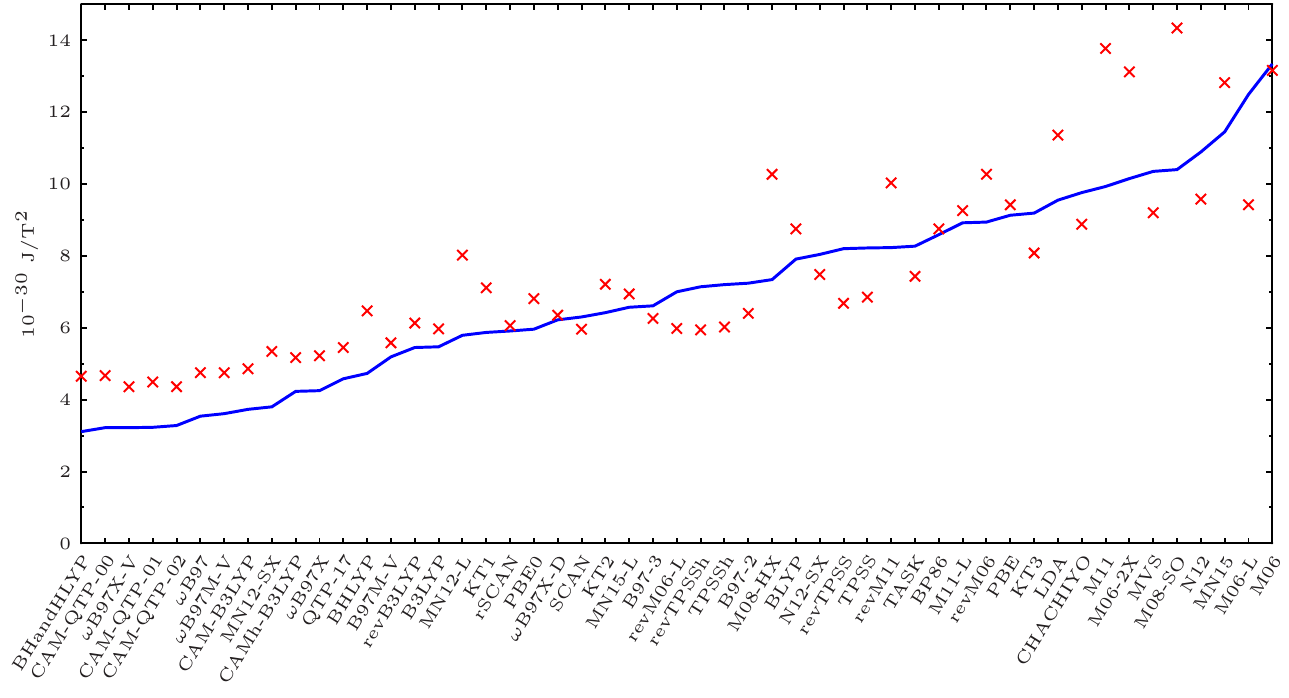}
  \caption{The mean absolute errors (blue solid line) as well as the
    errors' standard deviations (red crosses) of the magnetizabilities
    in $10^{-30}\, {\rm J}/{\rm T}^2$) of the 27 studied molecules
    obtained with the 51 functionals compared to the CCSD(T)
    reference.}
  \label{fig:MAE-and-STD}
\end{figure*}

We have calculated magnetizabilities for a series of small molecules
using both recently published density functionals, as well as older,
established density functionals. The accuracy of the magnetizabilities
predicted by the various density functional approximations has been
assessed by comparison to coupled-cluster calculations with singles
and doubles and perturbative triples [CCSD(T)] reported by
\citet{Lutnes2009_JCP_144104} Our results are summarized graphically
in \figref{MAE-and-STD}: the top functionals afford both small mean
absolute errors and standard deviations, but the same is not true for
all recently suggested functionals.

Numerical methods for calculating magnetizabilities based on
quadrature of the magnetizability density have been implemented.  We
have shown that this method allows studies of spatial contributions to
the magnetizabilities by visualization of the magnetizability density.
The method has been employed to calculate magnetizabilities from
magnetically induced current density susceptibilities, which were
obtained from \Turbomole{} calculations of nuclear magnetic shielding
constants. Thus, magnetizabilities can be calculated in this way with
\Turbomole{} even though analytical methods to calculate
magnetizabilities as the second derivative of the energy are not yet
available in this program.  Further information about spatial
contributions to the magnetizability could be obtained in the present
approach by studying atomic contributions and investigating the
positive and negative parts of the integrands separately in analogy to
our recent work on nuclear magnetic shieldings in \citeref{Jinger:20},
which may be studied in future work.

Our calculations show that the most accurate magnetizabilities (judged
by the smallest MAE) for the studied database are obtained with
BHandHLYP, which is an old global hybrid with 50\% HF exchange and
50\% B88 exchange. The calculations also show that the modern
range-separated functionals with 100\% long-range HF exchange
developed by Head-Gordon and co-workers and by Bartlett and co-workers
yield accurate magnetizabilities for the database. Calculations with
other range-separated functionals like CAM-B3LYP and CAMh-B3LYP as
well as with global hybrid functionals like QTP-17, BHLYP
a.k.a. BHandH, B3LYP and PBE0 yield relatively accurate
magnetizabilities for the studied molecules.  Meta-GGA functionals are
found to yield somewhat better magnetizabilities than GGA and LDA
functionals.

However, functionals developed by Truhlar and co-workers do not appear
to be well-aimed for calculations of magnetizabilities and other
magnetic properties that involve magnetically induced current
densities. Magnetizabilities calculated using the popular M06-2X
functional are found to be unreliable, and we do not recommend the use
of the M06-2X functional in calculations of nuclear magnetic
shieldings, magnetizabilities, ring-current strengths and other
magnetic properties that depend on magnetically induced current
density susceptibilities.  Previous studies have also suggested that
the M06-2X functional sometimes underestimates magnetizabilities and
ring-current strengths.\cite{Valiev2020_JPCC_21027,
  Valiev2018_JPCA_4756, Valiev2018_PCCP_17705} Revised versions of
Minnesota functionals have been studied in this work, and found to
yield somewhat more accurate magnetizabilities than the original
parameterizations. However, the revised versions also still appear on
the second half of the ranking list.

\begin{acknowledgement}

We thank Radovan Bast for help with the implementation of the
numerical integration in \Gimic{} using \Numgrid{}. This work has been
supported by the Academy of Finland (Suomen Akatemia) through project
numbers 311149 and 314821, by the Magnus Ehrnrooth Foundation, and by
The Swedish Cultural Foundation in Finland. We acknowledge
computational resources from the Finnish Grid and Cloud Infrastructure
(persistent identifier urn:nbn:fi:research-infras-2016072533) and CSC
-- IT Center for Science, Finland.
\end{acknowledgement}


\clearpage
\bibliography{literature,libxc,susi}

\clearpage
\begin{suppinfo}
\newcommand*\datacaption[1]{Magnetizabilities in units of $ 10^{-30}
  {\rm J}/{\rm T}^{2} $ for the #1 functionals in the aug-cc-pCVQZ
  basis set from calculations with \Turbomole{} and \Gimic{} compared
  to CCSD(T) data from \citeref{Lutnes2009_JCP_144104}.}

\newcommand*\diffcaption[1]{Deviations of the magnetizabilities
  computed with \Turbomole{} and \Gimic{} for the #1 functionals in
  the aug-cc-pCVQZ basis set from CCSD(T) data from
  \citeref{Lutnes2009_JCP_144104} in units of $ 10^{-30} {\rm J}/{\rm
    T}^{2} $.}

\newcommand*\pyscfdatacaption[1]{Magnetizabilities in units of $
  10^{-30} {\rm J}/{\rm T}^{2} $ for the #1 functionals in the
  aug-cc-pCVQZ basis set from calculations with \PySCF{} compared to
  CCSD(T) data from \citeref{Lutnes2009_JCP_144104}.}

\newcommand*\pyscfdiffcaption[1]{Deviations of the magnetizabilities
  computed with \PySCF{} for the #1 functionals in the aug-cc-pCVQZ
  basis set from CCSD(T) data from \citeref{Lutnes2009_JCP_144104} in
  units of $ 10^{-30} {\rm J}/{\rm T}^{2} $.}

Contents:
\begin{itemize}
\item Magnetically induced current-density susceptibilitise
\item Calculations on tetraoxa-isophlorin
\item \tabref{ST1}: magnetizabilities for B3LYP, B97-2, B97-3, B97M-V, and BHandHLYP
\item \tabref{ST2}: magnetizabilities for BHLYP, BLYP, BP86, and CAM-B3LYP
\item \tabref{ST3}: magnetizabilities for CAMh-B3LYP, CAM-QTP-00, and CAM-QTP-01
\item \tabref{ST4}: magnetizabilities for CAM-QTP-02, CHACHIYO, HF, KT1, KT2, and KT3
\item \tabref{ST5}: magnetizabilities for LDA, M06, M06-2X, M06-L, M08-HX, and M08-SO
\item \tabref{ST6}: magnetizabilities for M11, M11-L, MN12-L, MN12-SX, and MN15
\item \tabref{ST7}: magnetizabilities for MN15-L, MVS, N12, N12-SX, PBE, and PBE0
\item \tabref{ST8}: magnetizabilities for QTP-17, revB3LYP, revM06, and revM06-L
\item \tabref{ST9}: magnetizabilities for revM11, revTPSS, revTPSSh, rSCAN, and SCAN
\item \tabref{ST10}: magnetizabilities for TASK, TPSS, TPSSh, $\omega$B97, $\omega$B97M-V, and $\omega$B97X
\item \tabref{ST11}: magnetizabilities for $\omega$B97X-D, and $\omega$B97X-V
\item \tabref{ST1d}: magnetizability errors for B3LYP, B97-2, B97-3, B97M-V, and BHandHLYP
\item \tabref{ST2d}: magnetizability errors for BHLYP, BLYP, BP86, and CAM-B3LYP
\item \tabref{ST3d}: magnetizability errors for CAMh-B3LYP, CAM-QTP-00, and CAM-QTP-01
\item \tabref{ST4d}: magnetizability errors for CAM-QTP-02, CHACHIYO, HF, KT1, KT2, and KT3
\item \tabref{ST5d}: magnetizability errors for LDA, M06, M06-2X, M06-L, M08-HX, and M08-SO
\item \tabref{ST6d}: magnetizability errors for M11, M11-L, MN12-L, MN12-SX, and MN15
\item \tabref{ST7d}: magnetizability errors for MN15-L, MVS, N12, N12-SX, PBE, and PBE0
\item \tabref{ST8d}: magnetizability errors for QTP-17, revB3LYP, revM06, and revM06-L
\item \tabref{ST9d}: magnetizability errors for revM11, revTPSS, revTPSSh, rSCAN, and SCAN
\item \tabref{ST10d}: magnetizability errors for TASK, TPSS, TPSSh, $\omega$B97, $\omega$B97M-V, and $\omega$B97X
\item \tabref{ST11d}: magnetizability errors for $\omega$B97X-D, and $\omega$B97X-V
\item \tabref{comparison}: comparison of \Turbomole{} and \PySCF{} data.
\end{itemize}

\subsection*{Magnetically induced current-density susceptibilities}
\label{sec:MICD}

\begin{table*}
\begin{multline}
  {\cal J}^{B_\beta}_{\alpha} = \frac {\partial J_\alpha^{\bf B}} {\partial B_\beta} \Bigg|_{{\bf B}={\bf 0}}
  = \sum_{\mu\nu}D_{\mu\nu} \left[
\frac{\partial\chi^*_\mu(\mathbf{r})}{\partial B_\beta}
\frac{\partial \tilde h\mathbf{(r)}}{\partial m_{I_\alpha}}
\chi_\nu(\mathbf{r}) +
\chi^*_\mu(\mathbf{r})
\frac{\partial \tilde h\mathbf{(r)}}{\partial m_{I\alpha}}
\frac{\partial\chi_\nu(\mathbf{r})}{\partial B_\beta} \right.
\\
\left.
-\sum_{\gamma} \epsilon_{\alpha\beta\gamma}
\chi^*_\mu(\mathbf{r})
\frac{\partial^2 \tilde h\mathbf{(r)}}
{\partial m_{I_\alpha}\partial B_\gamma}\chi_\nu(\mathbf{r}) \right]_{{\bf B}={\bf 0}}
+
\sum_{\mu\nu} \left[
\frac{\partial D_{\mu\nu}}{\partial B_\beta}\chi^*_\mu(\mathbf{r})
\frac{\partial \tilde h\mathbf{(r)}}{\partial m_{I_\alpha}}\chi_\nu(\mathbf{r})\right]_{{\bf B}={\bf 0}}.
\label{eq:jexpression}
\end{multline}
\caption{The expression used to calculate the magnetically induced
current-density susceptibility (CDT).}
\label{tab:MCDS}
\end{table*}

The use of GIAOs eliminates the gauge origin ($\mathbf{R}_O$) from the
expression we use for calculating the CDT, which is given in
\eqref{jexpression} in \tabref{MCDS}. In the expression, $\mathbf{p}$
is the momentum operator, $m_{I_\alpha}$ are the Cartesian components
($\alpha$) of the magnetic moment of nucleus $I$, $B_\beta$ are the
Cartesian components ($\beta$) of the external magnetic field,
$\mathbf{D}$ is the density matrix in the atomic-orbital basis,
$[\partial \mathbf{D} /\partial \mathbf{B}]_{{\bf B}={\bf 0}}$ are the
magnetically perturbed density matrices,
$\epsilon_{\alpha\beta\gamma}$ is the Levi--Civita symbol, $\tilde
h\mathbf{(r)}$ denotes the magnetic interaction operator without the
$|\mathbf{r}-\mathbf{R}_I|^{-3}$ denominator with
\begin{equation}
  \frac {\partial \tilde h\mathbf{(r)}} {\partial \mathbf{m}_I} =
  (\mathbf{r}-\mathbf{R}_I)\times\mathbf{p}
  \label{eq:h_m}
\end{equation}
and
\begin{equation}
\frac{\partial^2\tilde h\mathbf{(r)}}{\partial \mathbf{m}_I\partial
  \mathbf{B}} = \frac{1}{2} [(\mathbf{r}-\mathbf{R}_O) \cdot
  (\mathbf{r}-\mathbf{R}_I)\mathbf{1} - (\mathbf{r}-\mathbf{R}_O)
  (\mathbf{r}-\mathbf{R}_I)],
\label{eq:h-mb}
\end{equation}
and $\mathbf{R}_I$ is the position of nucleus $I$. All terms that
contain the gauge origin $\mathbf{R}_O$ cancel in \eqref{jexpression},
making the CDT calculation independent of the gauge origin; this is
demonstrated in \figref{magnetizability-function2} for a different
choice of the gauge origin.  All terms containing the nuclear position
$\mathbf{R}_I$ also cancel, eliminating explicit references to the
nuclear coordinates.

\begin{figure*}
  \subfigure[]{\includegraphics[height=38mm]{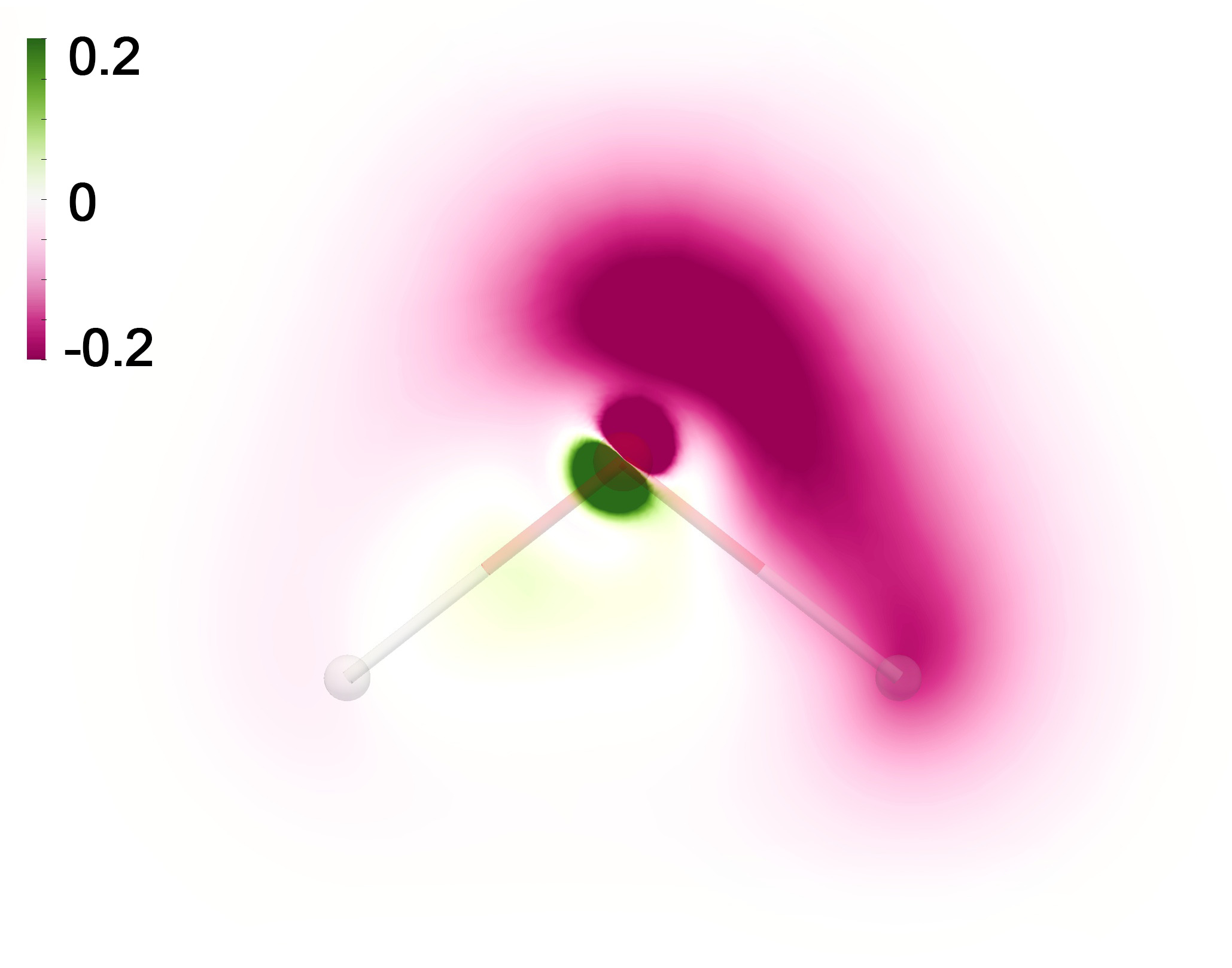} \label{fig:magnetizability-function2-h2o}}
  \subfigure[]{\includegraphics[height=38mm]{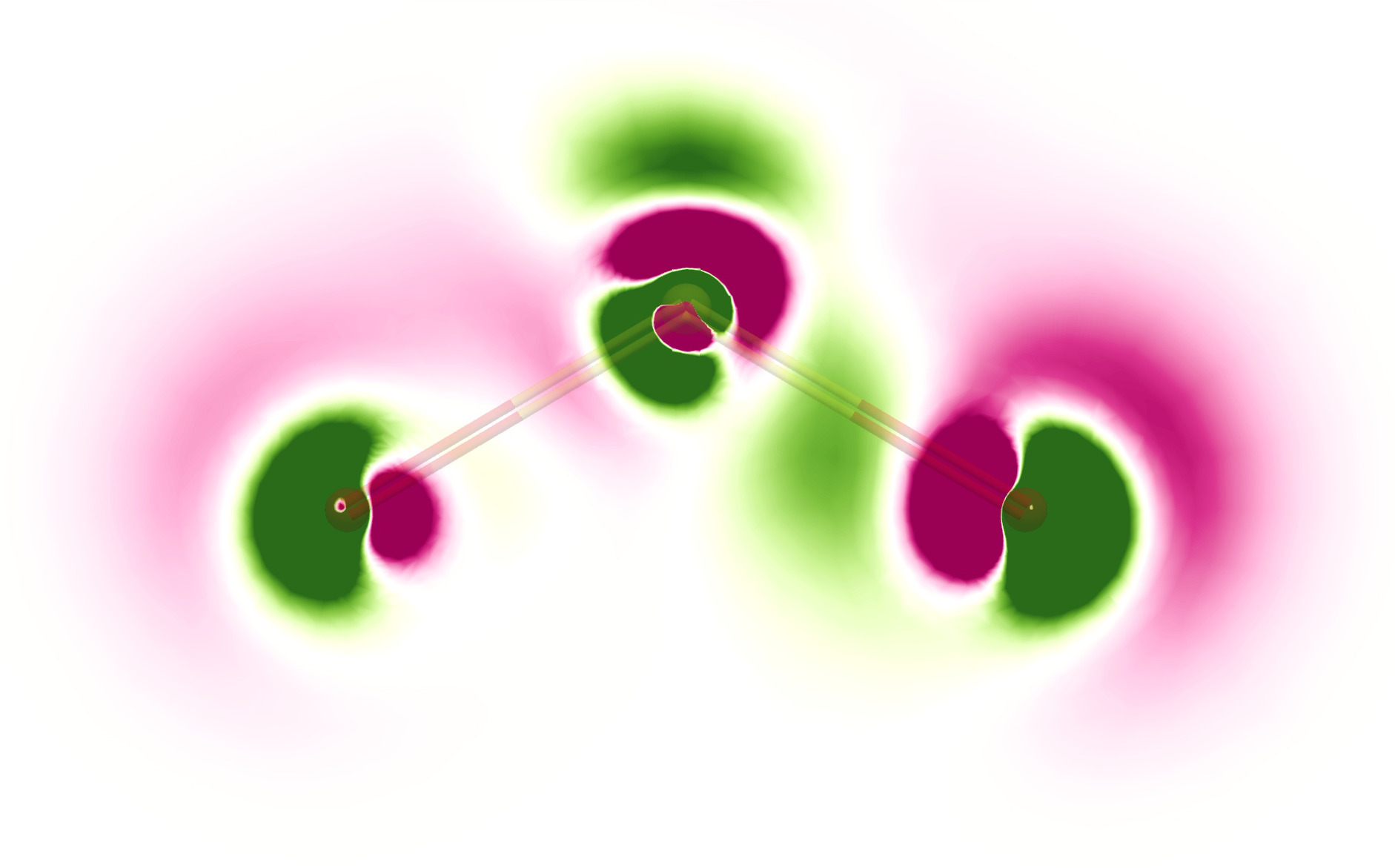} \label{fig:magnetizability-function2-so2}}
  \subfigure[]{\includegraphics[height=38mm]{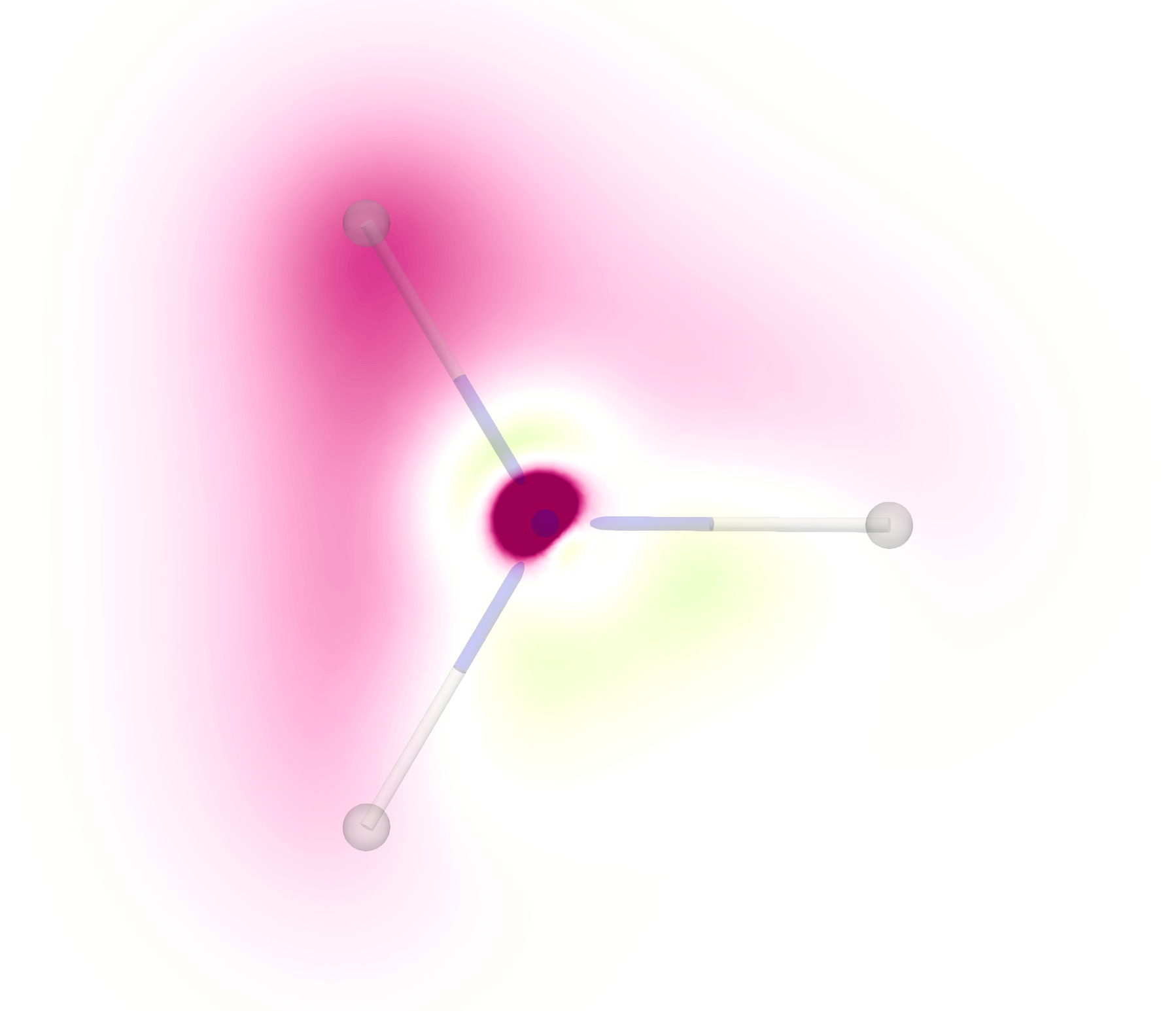} \label{fig:magnetizability-function2-nh3}}

  \caption{Visualization of the isotropic magnetizability density
    $\overline{\rho}^\xi({\bf r})$ shown in the molecular plane of
\ce{H2O} \ref{fig:magnetizability-function2-h2o}
and \ce{SO2} \ref{fig:magnetizability-function2-so2}
as well as in the plane formed by the hydrogen atoms of \ce{NH3}
\ref{fig:magnetizability-function2-nh3},
positioned \bohr{0.06} away from the \ce{N} atom towards the hydrogen atoms.
    Negative contributions are shown in pink, and positive
    ones in green. The gauge origin $\mathbf{R}_O$ is $(1,1,1)
    \bohr{}$.}
  \label{fig:magnetizability-function2}
\end{figure*}

\subsection*{Calculations on tetraoxa-isophlorin}

\citet{Valiev2017_CC_9866} found the isotropic magnetizability of
tetraoxa-isophlorin (molecule V in their work) to be 15.8 a.u.\ at the
LMP2/cc-pVDZ level of theory [local second-order M\o{}ller--Plesset
  perturbation theory], while B3LYP/def2-TZVP calculations yielded a
value of 65.9 a.u., which is over four times the LMP2 value.
Repeating the calculations of \citet{Valiev2017_CC_9866} with the
present approach using the def2-TZVP basis set, we obtained a
magnetizability of 65.2 a.u.\ at the B3LYP level, which agrees within
1\% with the value of \citet{Valiev2017_CC_9866}; this difference can
be tentatively attributed to the use of density fitting in the present
work. MN12-SX, which has no exact exchange in the long range, predicts
a susceptibility of 63.4 a.u., which is four times larger than the
LMP2/cc-pVDZ value and close to the B3LYP value.  Functionals with no
exact exchange like PBE yield even larger values, $>100$ a.u.  In
contrast, calculations at the CAM-B3LYP level with 65\% LR HF exchange
yield a magnetizability of 20.7 a.u., which agrees well with the LMP2
reference value. BHandHLYP contains 50\% LR HF exchange, and yields a
magnetizability of 23.7 a.u., which is also in qualitative agreement
with LMP2. Range-separated functionals with 100\% LR HF exchange like
$\omega$B97X and $\omega$B97 yield a magnetizability for
tetraoxa-isophlorin that is close to zero or even negative like the HF
value, which is $-11.6$ a.u.\cite{Valiev2017_CC_9866} The
B3LYP/def2-TZVP optimized geometry of \citeref{Valiev2017_CC_9866},
\textattachfile[color=0 0 1]{isophlorin.xyz}{attached here in xyz
  format}, was used in the calculations on tetraoxa-isophlorin.

\begin{table*}
\caption{\datacaption{B3LYP, B97-2, B97-3, B97M-V, and BHandHLYP}}
\begin{tabular}{lrrrrrr}
\hline
Molecule   &  B3LYP &  B97-2 &  B97-3 & B97M-V & BHandHLYP & CCSD(T) \tabularnewline
\hline
\hline
\ce{AlF}   & -396.5 & -393.4 & -392.6 & -396.1 & -395.6 & -394.5 \tabularnewline
\ce{C2H4}  & -336.7 & -334.3 & -334.6 & -334.8 & -343.0 & -345.6 \tabularnewline
\ce{C3H4}  & -463.1 & -462.6 & -464.1 & -460.6 & -468.8 & -478.9 \tabularnewline
\ce{CH2O}  & -114.9 & -116.6 & -115.4 & -132.8 & -123.8 & -127.4 \tabularnewline
\ce{CH3F}  & -312.4 & -312.2 & -313.3 & -309.4 & -314.9 & -315.7 \tabularnewline
\ce{CH4}   & -317.0 & -314.6 & -314.6 & -313.2 & -315.7 & -316.9 \tabularnewline
\ce{CO}    & -206.6 & -202.5 & -201.2 & -208.5 & -205.0 & -209.5 \tabularnewline
\hline
\end{tabular}
\label{tab:ST1}
\end{table*}

\begin{table*}
\caption{\datacaption{BHLYP, BLYP, BP86, and CAM-B3LYP}}
\begin{tabular}{lrrrrr}
\hline
Molecule   &  BHLYP &   BLYP &   BP86 & CAM-B3LYP & CCSD(T) \tabularnewline
\hline
\hline
\ce{AlF}   & -397.1 & -399.2 & -394.3 & -397.0 & -394.5 \tabularnewline
\ce{C2H4}  & -343.5 & -333.4 & -331.0 & -339.4 & -345.6 \tabularnewline
\ce{C3H4}  & -473.1 & -458.4 & -460.1 & -468.1 & -478.9 \tabularnewline
\ce{CH2O}  & -114.1 & -109.3 & -108.1 & -115.3 & -127.4 \tabularnewline
\ce{CH3F}  & -319.0 & -309.5 & -311.4 & -314.6 & -315.7 \tabularnewline
\ce{CH4}   & -325.2 & -318.1 & -318.5 & -320.0 & -316.9 \tabularnewline
\ce{CO}    & -205.9 & -209.1 & -205.2 & -208.4 & -209.5 \tabularnewline
\hline
\end{tabular}
\label{tab:ST2}
\end{table*}

\begin{table*}
\caption{\datacaption{CAMh-B3LYP, CAM-QTP-00, and CAM-QTP-01}}
\begin{tabular}{lrrrr}
\hline
Molecule   & CAMh-B3LYP & CAM-QTP-00 & CAM-QTP-01 & CCSD(T) \tabularnewline
\hline
\hline
\ce{AlF}   & -396.9 & -394.5 & -397.2 & -394.5 \tabularnewline
\ce{C2H4}  & -338.5 & -344.7 & -341.6 & -345.6 \tabularnewline
\ce{C3H4}  & -466.2 & -472.4 & -471.9 & -478.9 \tabularnewline
\ce{CH2O}  & -115.6 & -124.5 & -115.3 & -127.4 \tabularnewline
\ce{CH3F}  & -313.6 & -316.7 & -316.6 & -315.7 \tabularnewline
\ce{CH4}   & -318.6 & -317.5 & -322.7 & -316.9 \tabularnewline
\ce{CO}    & -207.9 & -205.2 & -209.4 & -209.5 \tabularnewline
\ce{FCCH}  & -441.2 & -444.8 & -443.5 & -441.6 \tabularnewline
\ce{FCN}   & -368.8 & -371.8 & -371.2 & -370.0 \tabularnewline
\ce{H2C2O} & -423.6 & -427.0 & -427.4 & -423.9 \tabularnewline
\ce{H2O}   & -237.2 & -234.0 & -238.0 & -235.1 \tabularnewline
\ce{H2S}   & -455.8 & -454.0 & -458.5 & -455.1 \tabularnewline
\ce{H4C2O} & -529.5 & -538.2 & -535.6 & -535.2 \tabularnewline
\ce{HCN}   & -271.2 & -273.9 & -273.9 & -271.8 \tabularnewline
\ce{HCP}   & -488.4 & -494.5 & -489.8 & -492.8 \tabularnewline
\ce{HF}    & -178.8 & -175.5 & -179.1 & -176.4 \tabularnewline
\ce{HFCO}  & -302.2 & -305.3 & -304.5 & -307.2 \tabularnewline
\ce{HOF}   & -232.5 & -238.7 & -235.6 & -235.4 \tabularnewline
\ce{LiF}   & -195.4 & -192.6 & -195.8 & -195.5 \tabularnewline
\ce{LiH}   & -129.8 & -124.9 & -128.6 & -127.2 \tabularnewline
\ce{N2}    & -203.6 & -202.3 & -205.6 & -205.2 \tabularnewline
\ce{N2O}   & -335.2 & -338.2 & -337.7 & -339.1 \tabularnewline
\ce{NH3}   & -292.0 & -289.8 & -293.9 & -290.3 \tabularnewline
\ce{O3}    &  250.1 &  373.1 &  283.2 &  121.5 \tabularnewline
\ce{OCS}   & -582.1 & -588.0 & -586.4 & -584.1 \tabularnewline
\ce{OF2}   & -237.7 & -255.0 & -244.9 & -247.1 \tabularnewline
\ce{PN}    & -295.9 & -297.6 & -300.3 & -308.2 \tabularnewline
\ce{SO2}   & -298.8 & -298.6 & -303.1 & -314.3 \tabularnewline
\hline
\end{tabular}
\label{tab:ST3}
\end{table*}

\begin{table*}
\caption{\datacaption{CAM-QTP-02, CHACHIYO, HF, KT1, KT2, and KT3}}
\begin{tabular}{lrrrrrrr}
\hline
Molecule   & CAM-QTP-02 & CHACHIYO &     HF &    KT1 &    KT2 &    KT3 & CCSD(T) \tabularnewline
\hline
\hline
\ce{AlF}   & -397.4 & -392.2 & -399.2 & -398.4 & -392.4 & -394.2 & -394.5 \tabularnewline
\ce{C2H4}  & -343.0 & -329.0 & -354.8 & -338.6 & -335.2 & -332.4 & -345.6 \tabularnewline
\ce{C3H4}  & -473.3 & -458.7 & -478.1 & -461.5 & -457.3 & -453.2 & -478.9 \tabularnewline
\ce{CH2O}  & -116.5 & -109.3 & -139.4 & -116.8 & -118.0 & -117.9 & -127.4 \tabularnewline
\ce{CH3F}  & -317.4 & -310.8 & -317.9 & -309.8 & -307.4 & -305.3 & -315.7 \tabularnewline
\ce{CH4}   & -323.3 & -315.5 & -313.6 & -320.7 & -316.0 & -311.8 & -316.9 \tabularnewline
\ce{CO}    & -209.3 & -202.6 & -204.5 & -214.0 & -209.1 & -206.1 & -209.5 \tabularnewline
\hline
\end{tabular}
\label{tab:ST4}
\end{table*}

\begin{table*}
\caption{\datacaption{LDA, M06, M06-2X, M06-L, M08-HX, and M08-SO}}
\begin{tabular}{lrrrrrrr}
\hline
Molecule   &    LDA &    M06 & M06-2X &  M06-L & M08-HX & M08-SO & CCSD(T) \tabularnewline
\hline
\hline
\ce{AlF}   & -395.8 & -387.1 & -392.9 & -382.7 & -397.2 & -395.8 & -394.5 \tabularnewline
\ce{C2H4}  & -331.1 & -332.4 & -329.9 & -327.1 & -331.2 & -334.3 & -345.6 \tabularnewline
\ce{C3H4}  & -464.4 & -465.3 & -461.4 & -461.5 & -462.7 & -463.7 & -478.9 \tabularnewline
\ce{CH2O}  &  -95.9 & -104.5 &  -94.1 & -123.6 &  -94.1 &  -89.4 & -127.4 \tabularnewline
\ce{CH3F}  & -315.4 & -313.4 & -317.2 & -311.1 & -319.5 & -319.0 & -315.7 \tabularnewline
\ce{CH4}   & -329.4 & -316.3 & -319.9 & -309.4 & -322.6 & -323.0 & -316.9 \tabularnewline
\ce{CO}    & -206.6 & -192.3 & -193.9 & -195.3 & -201.7 & -195.2 & -209.5 \tabularnewline
\hline
\end{tabular}
\label{tab:ST5}
\end{table*}

\begin{table*}
\caption{\datacaption{M11, M11-L, MN12-L, MN12-SX, and MN15}}
\begin{tabular}{lrrrrrr}
\hline
Molecule   &    M11 &  M11-L & MN12-L & MN12-SX &   MN15 & CCSD(T) \tabularnewline
\hline
\hline
\ce{AlF}   & -391.5 & -403.1 & -407.1 & -403.4 & -400.2 & -394.5 \tabularnewline
\ce{C2H4}  & -331.3 & -334.1 & -340.4 & -338.4 & -330.4 & -345.6 \tabularnewline
\ce{C3H4}  & -466.6 & -461.3 & -471.1 & -467.1 & -460.0 & -478.9 \tabularnewline
\ce{CH2O}  &  -89.0 & -135.4 & -145.0 & -128.4 &  -91.0 & -127.4 \tabularnewline
\ce{CH3F}  & -320.7 & -307.8 & -313.8 & -315.5 & -314.9 & -315.7 \tabularnewline
\ce{CH4}   & -323.7 & -312.0 & -315.6 & -317.7 & -319.4 & -316.9 \tabularnewline
\ce{CO}    & -199.2 & -203.4 & -211.4 & -207.8 & -194.3 & -209.5 \tabularnewline
\hline
\end{tabular}
\label{tab:ST6}
\end{table*}

\begin{table*}
\caption{\datacaption{MN15-L, MVS, N12, N12-SX, PBE, and PBE0}}
\begin{tabular}{lrrrrrrr}
\hline
Molecule   & MN15-L &    MVS &    N12 & N12-SX &    PBE &   PBE0 & CCSD(T) \tabularnewline
\hline
\hline
\ce{AlF}   & -410.2 & -384.3 & -394.1 & -394.7 & -396.9 & -394.3 & -394.5 \tabularnewline
\ce{C2H4}  & -343.2 & -319.8 & -331.1 & -333.8 & -330.8 & -335.3 & -345.6 \tabularnewline
\ce{C3H4}  & -470.4 & -459.6 & -456.3 & -463.8 & -459.5 & -465.1 & -478.9 \tabularnewline
\ce{CH2O}  & -142.4 & -123.5 & -112.9 & -113.3 & -104.9 & -112.8 & -127.4 \tabularnewline
\ce{CH3F}  & -314.1 & -309.1 & -307.8 & -313.8 & -311.2 & -314.4 & -315.7 \tabularnewline
\ce{CH4}   & -318.8 & -315.1 & -315.6 & -316.5 & -320.3 & -318.4 & -316.9 \tabularnewline
\ce{CO}    & -215.3 & -198.0 & -205.2 & -201.1 & -205.6 & -202.9 & -209.5 \tabularnewline
\hline
\end{tabular}
\label{tab:ST7}
\end{table*}

\begin{table*}
\caption{\datacaption{QTP-17, revB3LYP, revM06, and revM06-L}}
\begin{tabular}{lrrrrr}
\hline
Molecule   & QTP-17 & revB3LYP & revM06 & revM06-L & CCSD(T) \tabularnewline
\hline
\hline
\ce{AlF}   & -397.4 & -396.9 & -389.7 & -385.9 & -394.5 \tabularnewline
\ce{C2H4}  & -338.5 & -337.0 & -329.6 & -326.6 & -345.6 \tabularnewline
\ce{C3H4}  & -464.1 & -463.6 & -462.6 & -461.2 & -478.9 \tabularnewline
\ce{CH2O}  & -116.8 & -113.8 & -104.4 & -134.0 & -127.4 \tabularnewline
\ce{CH3F}  & -312.7 & -312.8 & -315.2 & -310.5 & -315.7 \tabularnewline
\ce{CH4}   & -316.9 & -318.3 & -318.4 & -312.0 & -316.9 \tabularnewline
\ce{CO}    & -207.1 & -206.8 & -197.2 & -206.4 & -209.5 \tabularnewline
\hline
\end{tabular}
\label{tab:ST8}
\end{table*}

\begin{table*}
\caption{\datacaption{revM11, revTPSS, revTPSSh, rSCAN, and SCAN}}
\begin{tabular}{lrrrrrr}
\hline
Molecule   & revM11 & revTPSS & revTPSSh &  rSCAN &   SCAN & CCSD(T) \tabularnewline
\hline
\hline
\ce{AlF}   & -393.5 & -393.2 & -392.7 & -395.0 & -392.0 & -394.5 \tabularnewline
\ce{C2H4}  & -333.0 & -332.0 & -333.7 & -332.3 & -333.0 & -345.6 \tabularnewline
\ce{C3H4}  & -467.4 & -457.7 & -460.2 & -463.4 & -462.4 & -478.9 \tabularnewline
\ce{CH2O}  &  -99.7 & -123.2 & -124.6 & -121.9 & -126.4 & -127.4 \tabularnewline
\ce{CH3F}  & -316.8 & -308.9 & -310.3 & -311.8 & -310.4 & -315.7 \tabularnewline
\ce{CH4}   & -320.1 & -308.8 & -309.3 & -318.5 & -314.4 & -316.9 \tabularnewline
\ce{CO}    & -201.9 & -204.9 & -204.1 & -206.7 & -206.9 & -209.5 \tabularnewline
\hline
\end{tabular}
\label{tab:ST9}
\end{table*}

\begin{table*}
\caption{\datacaption{TASK, TPSS, TPSSh, $\omega$B97, $\omega$B97M-V, and $\omega$B97X}}
\begin{tabular}{lrrrrrrr}
\hline
Molecule   &   TASK &   TPSS &  TPSSh & $\omega$B97 & $\omega$B97M-V & $\omega$B97X & CCSD(T) \tabularnewline
\hline
\hline
\ce{AlF}   & -383.3 & -393.9 & -393.3 & -397.0 & -400.1 & -395.5 & -394.5 \tabularnewline
\ce{C2H4}  & -325.4 & -332.1 & -333.8 & -338.1 & -340.5 & -337.5 & -345.6 \tabularnewline
\ce{C3H4}  & -460.8 & -458.5 & -460.9 & -471.5 & -469.9 & -469.6 & -478.9 \tabularnewline
\ce{CH2O}  & -132.9 & -120.0 & -121.7 & -118.9 & -115.8 & -116.4 & -127.4 \tabularnewline
\ce{CH3F}  & -309.3 & -309.5 & -310.9 & -315.3 & -315.0 & -314.9 & -315.7 \tabularnewline
\ce{CH4}   & -310.7 & -311.6 & -311.8 & -318.7 & -321.0 & -317.9 & -316.9 \tabularnewline
\ce{CO}    & -204.5 & -204.7 & -203.9 & -208.2 & -210.5 & -205.8 & -209.5 \tabularnewline
\hline
\end{tabular}
\label{tab:ST10}
\end{table*}

\begin{table*}
\caption{\datacaption{$\omega$B97X-D, and $\omega$B97X-V}}
\begin{tabular}{lrrr}
\hline
Molecule   & $\omega$B97X-D & $\omega$B97X-V & CCSD(T) \tabularnewline
\hline
\hline
\ce{AlF}   & -393.1 & -396.5 & -394.5 \tabularnewline
\ce{C2H4}  & -335.5 & -338.9 & -345.6 \tabularnewline
\ce{C3H4}  & -466.9 & -469.9 & -478.9 \tabularnewline
\ce{CH2O}  & -114.7 & -118.6 & -127.4 \tabularnewline
\ce{CH3F}  & -314.1 & -314.9 & -315.7 \tabularnewline
\ce{CH4}   & -316.1 & -317.7 & -316.9 \tabularnewline
\ce{CO}    & -202.4 & -207.4 & -209.5 \tabularnewline
\ce{FCCH}  & -439.8 & -442.4 & -441.6 \tabularnewline
\ce{FCN}   & -366.2 & -370.1 & -370.0 \tabularnewline
\ce{H2C2O} & -422.1 & -425.3 & -423.9 \tabularnewline
\ce{H2O}   & -233.8 & -236.0 & -235.1 \tabularnewline
\ce{H2S}   & -452.9 & -454.2 & -455.1 \tabularnewline
\ce{H4C2O} & -530.1 & -532.8 & -535.2 \tabularnewline
\ce{HCN}   & -266.2 & -270.7 & -271.8 \tabularnewline
\ce{HCP}   & -482.7 & -486.5 & -492.8 \tabularnewline
\ce{HF}    & -175.9 & -178.1 & -176.4 \tabularnewline
\ce{HFCO}  & -297.9 & -303.1 & -307.2 \tabularnewline
\ce{HOF}   & -232.0 & -234.7 & -235.4 \tabularnewline
\ce{LiF}   & -195.6 & -196.4 & -195.5 \tabularnewline
\ce{LiH}   & -131.5 & -127.4 & -127.2 \tabularnewline
\ce{N2}    & -197.7 & -202.9 & -205.2 \tabularnewline
\ce{N2O}   & -334.3 & -336.8 & -339.1 \tabularnewline
\ce{NH3}   & -288.7 & -290.5 & -290.3 \tabularnewline
\ce{O3}    &  267.2 &  251.2 &  121.5 \tabularnewline
\ce{OCS}   & -579.6 & -584.7 & -584.1 \tabularnewline
\ce{OF2}   & -238.0 & -243.8 & -247.1 \tabularnewline
\ce{PN}    & -286.5 & -296.0 & -308.2 \tabularnewline
\ce{SO2}   & -290.6 & -299.8 & -314.3 \tabularnewline
\hline
\end{tabular}
\label{tab:ST11}
\end{table*}

\begin{table*}
\caption{\diffcaption{B3LYP, B97-2, B97-3, B97M-V, and BHandHLYP}}
\begin{threeparttable}
\begin{tabular}{lrrrrr}
\hline
Molecule   &  B3LYP &  B97-2 &  B97-3 & B97M-V & BHandHLYP \tabularnewline
\hline
\hline
\ce{AlF}   &   -2.0 &    1.1 &    1.9 &   -1.6 &   -1.1 \tabularnewline
\ce{C2H4}  &    8.9 &   11.3 &   11.0 &   10.8 &    2.6 \tabularnewline
\ce{C3H4}  &   15.8 &   16.3 &   14.8 &   18.3 &   10.1 \tabularnewline
\ce{CH2O}  &   12.5 &   10.8 &   12.0 &   -5.4 &    3.6 \tabularnewline
\ce{CH3F}  &    3.3 &    3.5 &    2.4 &    6.3 &    0.8 \tabularnewline
\ce{CH4}   &   -0.1 &    2.3 &    2.3 &    3.7 &    1.2 \tabularnewline
\ce{CO}    &    2.9 &    7.0 &    8.3 &    1.0 &    4.5 \tabularnewline
\hline
\end{tabular}
\begin{tablenotes}
  \item [$^*$] Statistics for the mean absolute error (MAE), mean error (ME) and the standard deviation (STD) exclude \ce{O3}.
\end{tablenotes}
\end{threeparttable}
\label{tab:ST1d}
\end{table*}

\begin{table*}
\caption{\diffcaption{BHLYP, BLYP, BP86, and CAM-B3LYP}}
\begin{threeparttable}
\begin{tabular}{lrrrr}
\hline
Molecule   &  BHLYP &   BLYP &   BP86 & CAM-B3LYP \tabularnewline
\hline
\hline
\ce{AlF}   &   -2.6 &   -4.7 &    0.2 &   -2.5 \tabularnewline
\ce{C2H4}  &    2.1 &   12.2 &   14.6 &    6.2 \tabularnewline
\ce{C3H4}  &    5.8 &   20.5 &   18.8 &   10.8 \tabularnewline
\ce{CH2O}  &   13.3 &   18.1 &   19.3 &   12.1 \tabularnewline
\ce{CH3F}  &   -3.3 &    6.2 &    4.3 &    1.1 \tabularnewline
\ce{CH4}   &   -8.3 &   -1.2 &   -1.6 &   -3.1 \tabularnewline
\ce{CO}    &    3.6 &    0.4 &    4.3 &    1.1 \tabularnewline
\hline
\end{tabular}
\begin{tablenotes}
  \item [$^*$] Statistics for the mean absolute error (MAE), mean error (ME) and the standard deviation (STD) exclude \ce{O3}.
\end{tablenotes}
\end{threeparttable}
\label{tab:ST2d}
\end{table*}

\begin{table*}
\caption{\diffcaption{CAMh-B3LYP, CAM-QTP-00, and CAM-QTP-01}}
\begin{threeparttable}
\begin{tabular}{lrrr}
\hline
Molecule   & CAMh-B3LYP & CAM-QTP-00 & CAM-QTP-01 \tabularnewline
\hline
\hline
\ce{AlF}   &   -2.4 &   -0.0 &   -2.7 \tabularnewline
\ce{C2H4}  &    7.1 &    0.9 &    4.0 \tabularnewline
\ce{C3H4}  &   12.7 &    6.5 &    7.0 \tabularnewline
\ce{CH2O}  &   11.8 &    2.9 &   12.1 \tabularnewline
\ce{CH3F}  &    2.1 &   -1.0 &   -0.9 \tabularnewline
\ce{CH4}   &   -1.7 &   -0.6 &   -5.8 \tabularnewline
\ce{CO}    &    1.6 &    4.3 &    0.1 \tabularnewline
\ce{FCCH}  &    0.4 &   -3.2 &   -1.9 \tabularnewline
\ce{FCN}   &    1.2 &   -1.8 &   -1.2 \tabularnewline
\ce{H2C2O} &    0.3 &   -3.1 &   -3.5 \tabularnewline
\ce{H2O}   &   -2.1 &    1.1 &   -2.9 \tabularnewline
\ce{H2S}   &   -0.7 &    1.1 &   -3.4 \tabularnewline
\ce{H4C2O} &    5.7 &   -3.0 &   -0.4 \tabularnewline
\ce{HCN}   &    0.6 &   -2.1 &   -2.1 \tabularnewline
\ce{HCP}   &    4.4 &   -1.7 &    3.0 \tabularnewline
\ce{HF}    &   -2.4 &    0.9 &   -2.7 \tabularnewline
\ce{HFCO}  &    5.0 &    1.9 &    2.7 \tabularnewline
\ce{HOF}   &    2.9 &   -3.3 &   -0.2 \tabularnewline
\ce{LiF}   &    0.1 &    2.9 &   -0.3 \tabularnewline
\ce{LiH}   &   -2.6 &    2.3 &   -1.4 \tabularnewline
\ce{N2}    &    1.6 &    2.9 &   -0.4 \tabularnewline
\ce{N2O}   &    3.9 &    0.9 &    1.4 \tabularnewline
\ce{NH3}   &   -1.7 &    0.5 &   -3.6 \tabularnewline
\ce{O3}    &  128.6 &  251.6 &  161.7 \tabularnewline
\ce{OCS}   &    2.0 &   -3.9 &   -2.3 \tabularnewline
\ce{OF2}   &    9.4 &   -7.9 &    2.2 \tabularnewline
\ce{PN}    &   12.3 &   10.6 &    7.9 \tabularnewline
\ce{SO2}   &   15.5 &   15.7 &   11.2 \tabularnewline
\hline
MAE$^*$    &    4.2 &    3.2 &    3.2 \tabularnewline
ME$^*$     &    3.2 &    0.9 &    0.6 \tabularnewline
STD$^*$    &    5.2 &    4.7 &    4.5 \tabularnewline
\hline
\end{tabular}
\begin{tablenotes}
  \item [$^*$] Statistics for the mean absolute error (MAE), mean error (ME) and the standard deviation (STD) exclude \ce{O3}.
\end{tablenotes}
\end{threeparttable}
\label{tab:ST3d}
\end{table*}

\begin{table*}
\caption{\diffcaption{CAM-QTP-02, CHACHIYO, HF, KT1, KT2, and KT3}}
\begin{threeparttable}
\begin{tabular}{lrrrrrr}
\hline
Molecule   & CAM-QTP-02 & CHACHIYO &     HF &    KT1 &    KT2 &    KT3 \tabularnewline
\hline
\hline
\ce{AlF}   &   -2.9 &    2.3 &   -4.7 &   -3.9 &    2.1 &    0.3 \tabularnewline
\ce{C2H4}  &    2.6 &   16.6 &   -9.2 &    7.0 &   10.4 &   13.2 \tabularnewline
\ce{C3H4}  &    5.6 &   20.2 &    0.8 &   17.4 &   21.6 &   25.7 \tabularnewline
\ce{CH2O}  &   10.9 &   18.1 &  -12.0 &   10.6 &    9.4 &    9.5 \tabularnewline
\ce{CH3F}  &   -1.7 &    4.9 &   -2.2 &    5.9 &    8.3 &   10.4 \tabularnewline
\ce{CH4}   &   -6.4 &    1.4 &    3.3 &   -3.8 &    0.9 &    5.1 \tabularnewline
\ce{CO}    &    0.2 &    6.9 &    5.0 &   -4.5 &    0.4 &    3.4 \tabularnewline
\hline
\end{tabular}
\begin{tablenotes}
  \item [$^*$] Statistics for the mean absolute error (MAE), mean error (ME) and the standard deviation (STD) exclude \ce{O3}.
\end{tablenotes}
\end{threeparttable}
\label{tab:ST4d}
\end{table*}

\begin{table*}
\caption{\diffcaption{LDA, M06, M06-2X, M06-L, M08-HX, and M08-SO}}
\begin{threeparttable}
\begin{tabular}{lrrrrrr}
\hline
Molecule   &    LDA &    M06 & M06-2X &  M06-L & M08-HX & M08-SO \tabularnewline
\hline
\hline
\ce{AlF}   &   -1.3 &    7.4 &    1.6 &   11.8 &   -2.7 &   -1.3 \tabularnewline
\ce{C2H4}  &   14.5 &   13.2 &   15.7 &   18.5 &   14.4 &   11.3 \tabularnewline
\ce{C3H4}  &   14.5 &   13.6 &   17.5 &   17.4 &   16.2 &   15.2 \tabularnewline
\ce{CH2O}  &   31.5 &   22.9 &   33.3 &    3.8 &   33.3 &   38.0 \tabularnewline
\ce{CH3F}  &    0.3 &    2.3 &   -1.5 &    4.6 &   -3.8 &   -3.3 \tabularnewline
\ce{CH4}   &  -12.5 &    0.6 &   -3.0 &    7.5 &   -5.7 &   -6.1 \tabularnewline
\ce{CO}    &    2.9 &   17.2 &   15.6 &   14.2 &    7.8 &   14.3 \tabularnewline
\hline
\end{tabular}
\begin{tablenotes}
  \item [$^*$] Statistics for the mean absolute error (MAE), mean error (ME) and the standard deviation (STD) exclude \ce{O3}.
\end{tablenotes}
\end{threeparttable}
\label{tab:ST5d}
\end{table*}

\begin{table*}
\caption{\diffcaption{M11, M11-L, MN12-L, MN12-SX, and MN15}}
\begin{threeparttable}
\begin{tabular}{lrrrrr}
\hline
Molecule   &    M11 &  M11-L & MN12-L & MN12-SX &   MN15 \tabularnewline
\hline
\hline
\ce{AlF}   &    3.0 &   -8.6 &  -12.6 &   -8.9 &   -5.7 \tabularnewline
\ce{C2H4}  &   14.3 &   11.5 &    5.2 &    7.2 &   15.2 \tabularnewline
\ce{C3H4}  &   12.3 &   17.6 &    7.8 &   11.8 &   18.9 \tabularnewline
\ce{CH2O}  &   38.4 &   -8.0 &  -17.6 &   -1.0 &   36.4 \tabularnewline
\ce{CH3F}  &   -5.0 &    7.9 &    1.9 &    0.2 &    0.8 \tabularnewline
\ce{CH4}   &   -6.8 &    4.9 &    1.3 &   -0.8 &   -2.5 \tabularnewline
\ce{CO}    &   10.3 &    6.1 &   -1.9 &    1.7 &   15.2 \tabularnewline
\hline
\end{tabular}
\begin{tablenotes}
  \item [$^*$] Statistics for the mean absolute error (MAE), mean error (ME) and the standard deviation (STD) exclude \ce{O3}.
\end{tablenotes}
\end{threeparttable}
\label{tab:ST6d}
\end{table*}

\begin{table*}
\caption{\diffcaption{MN15-L, MVS, N12, N12-SX, PBE, and PBE0}}
\begin{threeparttable}
\begin{tabular}{lrrrrrr}
\hline
Molecule   & MN15-L &    MVS &    N12 & N12-SX &    PBE &   PBE0 \tabularnewline
\hline
\hline
\ce{AlF}   &  -15.7 &   10.2 &    0.4 &   -0.2 &   -2.4 &    0.2 \tabularnewline
\ce{C2H4}  &    2.4 &   25.8 &   14.5 &   11.8 &   14.8 &   10.3 \tabularnewline
\ce{C3H4}  &    8.5 &   19.3 &   22.6 &   15.1 &   19.4 &   13.8 \tabularnewline
\ce{CH2O}  &  -15.0 &    3.9 &   14.5 &   14.1 &   22.5 &   14.6 \tabularnewline
\ce{CH3F}  &    1.6 &    6.6 &    7.9 &    1.9 &    4.5 &    1.3 \tabularnewline
\ce{CH4}   &   -1.9 &    1.8 &    1.3 &    0.4 &   -3.4 &   -1.5 \tabularnewline
\ce{CO}    &   -5.8 &   11.5 &    4.3 &    8.4 &    3.9 &    6.6 \tabularnewline
\hline
\end{tabular}
\begin{tablenotes}
  \item [$^*$] Statistics for the mean absolute error (MAE), mean error (ME) and the standard deviation (STD) exclude \ce{O3}.
\end{tablenotes}
\end{threeparttable}
\label{tab:ST7d}
\end{table*}

\begin{table*}
\caption{\diffcaption{QTP-17, revB3LYP, revM06, and revM06-L}}
\begin{threeparttable}
\begin{tabular}{lrrrr}
\hline
Molecule   & QTP-17 & revB3LYP & revM06 & revM06-L \tabularnewline
\hline
\hline
\ce{AlF}   &   -2.9 &   -2.4 &    4.8 &    8.6 \tabularnewline
\ce{C2H4}  &    7.1 &    8.6 &   16.0 &   19.0 \tabularnewline
\ce{C3H4}  &   14.8 &   15.3 &   16.3 &   17.7 \tabularnewline
\ce{CH2O}  &   10.6 &   13.6 &   23.0 &   -6.6 \tabularnewline
\ce{CH3F}  &    3.0 &    2.9 &    0.5 &    5.2 \tabularnewline
\ce{CH4}   &   -0.0 &   -1.4 &   -1.5 &    4.9 \tabularnewline
\ce{CO}    &    2.4 &    2.7 &   12.3 &    3.1 \tabularnewline
\hline
\end{tabular}
\begin{tablenotes}
  \item [$^*$] Statistics for the mean absolute error (MAE), mean error (ME) and the standard deviation (STD) exclude \ce{O3}.
\end{tablenotes}
\end{threeparttable}
\label{tab:ST8d}
\end{table*}

\begin{table*}
\caption{\diffcaption{revM11, revTPSS, revTPSSh, rSCAN, and SCAN}}
\begin{threeparttable}
\begin{tabular}{lrrrrr}
\hline
Molecule   & revM11 & revTPSS & revTPSSh &  rSCAN &   SCAN \tabularnewline
\hline
\hline
\ce{AlF}   &    1.0 &    1.3 &    1.8 &   -0.5 &    2.5 \tabularnewline
\ce{C2H4}  &   12.6 &   13.6 &   11.9 &   13.3 &   12.6 \tabularnewline
\ce{C3H4}  &   11.5 &   21.2 &   18.7 &   15.5 &   16.5 \tabularnewline
\ce{CH2O}  &   27.7 &    4.2 &    2.8 &    5.5 &    1.0 \tabularnewline
\ce{CH3F}  &   -1.1 &    6.8 &    5.4 &    3.9 &    5.3 \tabularnewline
\ce{CH4}   &   -3.2 &    8.1 &    7.6 &   -1.6 &    2.5 \tabularnewline
\ce{CO}    &    7.6 &    4.6 &    5.4 &    2.8 &    2.6 \tabularnewline
\hline
\end{tabular}
\begin{tablenotes}
  \item [$^*$] Statistics for the mean absolute error (MAE), mean error (ME) and the standard deviation (STD) exclude \ce{O3}.
\end{tablenotes}
\end{threeparttable}
\label{tab:ST9d}
\end{table*}

\begin{table*}
\caption{\diffcaption{TASK, TPSS, TPSSh, $\omega$B97, $\omega$B97M-V, and $\omega$B97X}}
\begin{threeparttable}
\begin{tabular}{lrrrrrr}
\hline
Molecule   &   TASK &   TPSS &  TPSSh & $\omega$B97 & $\omega$B97M-V & $\omega$B97X \tabularnewline
\hline
\hline
\ce{AlF}   &   11.2 &    0.6 &    1.2 &   -2.5 &   -5.6 &   -1.0 \tabularnewline
\ce{C2H4}  &   20.2 &   13.5 &   11.8 &    7.5 &    5.1 &    8.1 \tabularnewline
\ce{C3H4}  &   18.1 &   20.4 &   18.0 &    7.4 &    9.0 &    9.3 \tabularnewline
\ce{CH2O}  &   -5.5 &    7.4 &    5.7 &    8.5 &   11.6 &   11.0 \tabularnewline
\ce{CH3F}  &    6.4 &    6.2 &    4.8 &    0.4 &    0.7 &    0.8 \tabularnewline
\ce{CH4}   &    6.2 &    5.3 &    5.1 &   -1.8 &   -4.1 &   -1.0 \tabularnewline
\ce{CO}    &    5.0 &    4.8 &    5.6 &    1.3 &   -1.0 &    3.7 \tabularnewline
\hline
\end{tabular}
\begin{tablenotes}
  \item [$^*$] Statistics for the mean absolute error (MAE), mean error (ME) and the standard deviation (STD) exclude \ce{O3}.
\end{tablenotes}
\end{threeparttable}
\label{tab:ST10d}
\end{table*}

\begin{table*}
\caption{\diffcaption{$\omega$B97X-D, and $\omega$B97X-V}}
\begin{threeparttable}
\begin{tabular}{lrr}
\hline
Molecule   & $\omega$B97X-D & $\omega$B97X-V \tabularnewline
\hline
\hline
\ce{AlF}   &    1.4 &   -2.0 \tabularnewline
\ce{C2H4}  &   10.1 &    6.7 \tabularnewline
\ce{C3H4}  &   12.0 &    9.0 \tabularnewline
\ce{CH2O}  &   12.7 &    8.8 \tabularnewline
\ce{CH3F}  &    1.6 &    0.8 \tabularnewline
\ce{CH4}   &    0.8 &   -0.8 \tabularnewline
\ce{CO}    &    7.1 &    2.1 \tabularnewline
\ce{FCCH}  &    1.8 &   -0.8 \tabularnewline
\ce{FCN}   &    3.8 &   -0.1 \tabularnewline
\ce{H2C2O} &    1.8 &   -1.4 \tabularnewline
\ce{H2O}   &    1.3 &   -0.9 \tabularnewline
\ce{H2S}   &    2.2 &    0.9 \tabularnewline
\ce{H4C2O} &    5.1 &    2.4 \tabularnewline
\ce{HCN}   &    5.6 &    1.1 \tabularnewline
\ce{HCP}   &   10.1 &    6.3 \tabularnewline
\ce{HF}    &    0.5 &   -1.7 \tabularnewline
\ce{HFCO}  &    9.3 &    4.1 \tabularnewline
\ce{HOF}   &    3.4 &    0.7 \tabularnewline
\ce{LiF}   &   -0.1 &   -0.9 \tabularnewline
\ce{LiH}   &   -4.3 &   -0.2 \tabularnewline
\ce{N2}    &    7.5 &    2.3 \tabularnewline
\ce{N2O}   &    4.8 &    2.3 \tabularnewline
\ce{NH3}   &    1.6 &   -0.2 \tabularnewline
\ce{O3}    &  145.7 &  129.7 \tabularnewline
\ce{OCS}   &    4.5 &   -0.6 \tabularnewline
\ce{OF2}   &    9.1 &    3.3 \tabularnewline
\ce{PN}    &   21.7 &   12.2 \tabularnewline
\ce{SO2}   &   23.7 &   14.5 \tabularnewline
\hline
MAE$^*$    &    6.2 &    3.2 \tabularnewline
ME$^*$     &    5.9 &    2.5 \tabularnewline
STD$^*$    &    6.3 &    4.4 \tabularnewline
\hline
\end{tabular}
\begin{tablenotes}
  \item [$^*$] Statistics for the mean absolute error (MAE), mean error (ME) and the standard deviation (STD) exclude \ce{O3}.
\end{tablenotes}
\end{threeparttable}
\label{tab:ST11d}
\end{table*}

\begin{table*}
\caption{Comparison of the magnetizabilities in $ 10^{-30} {\rm J}/{\rm T}^{2} $
  calculated with \Turbomole{} (\textsc{TM}) /\Gimic{} employing the
  resolution of the identity approximation, and \PySCF{} employing exact integrals at the BP86/aug-cc-pCVQZ and
  B3LYP/aug-cc-pCVQZ levels of theory. The \PySCF{} data is in full agreement with that from \Gaussian{}.}
\begin{tabular}{c|rrr|rrr}
\hline
Molecule   & \multicolumn{3}{c}{BP86}
           & \multicolumn{3}{c}{B3LYP} \tabularnewline
           &  \textsc{Tm}/\Gimic{}  & \PySCF{} & difference  & \textsc{Tm}/\Gimic{} & \PySCF{} & difference \tabularnewline
\hline
\hline
\ce{AlF}   & -394.3 & -394.4 &    0.2 & -396.5 & -396.6 &    0.1 \tabularnewline
\ce{C2H4}  & -331.0 & -330.9 &   -0.0 & -336.7 & -336.7 &   -0.0 \tabularnewline
\ce{C3H4}  & -460.1 & -460.1 &   -0.0 & -463.1 & -463.0 &   -0.0 \tabularnewline
\ce{CH2O}  & -108.1 & -108.1 &    0.0 & -114.9 & -114.9 &    0.0 \tabularnewline
\ce{CH3F}  & -311.4 & -311.4 &    0.0 & -312.4 & -312.4 &    0.0 \tabularnewline
\ce{CH4}   & -318.5 & -318.6 &    0.1 & -317.0 & -317.1 &    0.1 \tabularnewline
\ce{CO}    & -205.2 & -205.2 &    0.0 & -206.6 & -206.6 &    0.0 \tabularnewline
\ce{FCCH}  & -437.9 & -437.9 &   -0.0 & -440.1 & -440.0 &   -0.0 \tabularnewline
\ce{FCN}   & -365.0 & -365.0 &    0.0 & -367.4 & -367.4 &   -0.0 \tabularnewline
\ce{H2C2O} & -422.0 & -422.1 &    0.1 & -422.1 & -422.2 &    0.1 \tabularnewline
\ce{H2O}   & -237.5 & -237.5 &    0.1 & -236.7 & -236.7 &    0.0 \tabularnewline
\ce{H2S}   & -456.6 & -456.7 &    0.2 & -455.1 & -455.3 &    0.2 \tabularnewline
\ce{H4C2O} & -523.5 & -523.5 &    0.0 & -526.9 & -526.9 &   -0.0 \tabularnewline
\ce{HCN}   & -264.4 & -264.4 &    0.0 & -269.4 & -269.4 &    0.0 \tabularnewline
\ce{HCP}   & -479.2 & -479.3 &    0.0 & -487.4 & -487.4 &   -0.0 \tabularnewline
\ce{HF}    & -179.3 & -179.3 &    0.0 & -178.4 & -178.4 &    0.0 \tabularnewline
\ce{HFCO}  & -296.5 & -296.5 &    0.0 & -300.5 & -300.5 &    0.0 \tabularnewline
\ce{HOF}   & -227.8 & -227.7 &   -0.0 & -231.1 & -231.1 &   -0.0 \tabularnewline
\ce{LiF}   & -197.0 & -197.1 &    0.1 & -194.7 & -194.8 &    0.1 \tabularnewline
\ce{LiH}   & -133.2 & -133.2 &   -0.0 & -130.8 & -130.7 &   -0.0 \tabularnewline
\ce{N2}    & -199.6 & -199.5 &   -0.1 & -202.0 & -201.9 &   -0.1 \tabularnewline
\ce{N2O}   & -332.6 & -332.7 &    0.1 & -333.8 & -334.0 &    0.1 \tabularnewline
\ce{NH3}   & -291.9 & -292.0 &    0.1 & -291.2 & -291.3 &    0.1 \tabularnewline
\ce{O3}    &  180.9 &  181.2 &   -0.3 &  238.7 &  239.0 &   -0.3 \tabularnewline
\ce{OCS}   & -575.0 & -575.1 &    0.1 & -579.6 & -579.7 &    0.1 \tabularnewline
\ce{OF2}   & -222.1 & -221.8 &   -0.2 & -234.1 & -233.9 &   -0.2 \tabularnewline
\ce{PN}    & -284.7 & -284.3 &   -0.4 & -292.2 & -291.8 &   -0.4 \tabularnewline
\ce{SO2}   & -292.7 & -292.3 &   -0.5 & -296.1 & -295.6 &   -0.4 \tabularnewline
\hline
\end{tabular}
\label{tab:comparison}
\end{table*}

\end{suppinfo}

\end{document}